\documentclass{aa520}
\usepackage{graphicx}
\usepackage{rotating}
\usepackage{txfonts}
\usepackage{pstricks}

\newcommand{\Ci}{{\it a\/}}
\newcommand{\D}[1]{{\rm d}#1}

\newcommand{\rsn}{\nu_0}
\newcommand{\rsl}{\lambda_0}

\newcommand{\Eq}[2]{\parbox{#1}{\vspace{-12pt}\begin{equation}#2\end{equation}}}
\newcommand{\Ma}[1]{\Eq{120pt}{#1}}
\newcommand{\Md}[1]{\Eq{130pt}{#1}}
\newcommand{\Mc}[1]{\Eq{100pt}{#1}}
\newcommand{\Tq}[2]{\parbox{#1}{\vspace{-12pt}\hbox{#2}}}
\newcommand{\TT}[1]{\Tq{40pt}{{\tt #1}}}
\newcommand{\Wh}{\Tq{25pt}{where}}
\newcommand{\Wi}{\Tq{15pt}{with}}

\newcommand{\vS}{S}
\newcommand{\vw}{w}
\newcommand{\vX}{X}
\newcommand{\vP}{P}
\newcommand{\vv}{\varv}
\newcommand{\eq}{\hbox{\hspace{0.6em}=\hspace{0.6em}}}

\newcommand{\sub}[1]{_\mathrm{#1}}
\newcommand{\HI}{H{\small I}}

\newcommand{\Dfrac}[2]{\frac{\D{#1}}{\D{#2}}}
\newcommand{\Dfracr}[2]{\left. \Dfrac{#1}{#2} \right|_{\rm r} \,}

\newcommand{\Ddiv}[2]{\D{#1}/\D{#2}}
\newcommand{\Ddivr}[2]{\left. \Ddiv{#1}{#2} \right|_{\rm r}}

\newcommand{\keyw}[1]{\hbox{{\tt #1}}}

\newcommand{\keyi}[2]{\hbox{{\tt #1\hspace{1pt}}{$#2$}\/}}
\newcommand{\CRPIX}[1]{\keyi{CRPIX}{#1}}
\newcommand{\CDELT}[1]{\keyi{CDELT}{#1}}
\newcommand{\CRVAL}[1]{\keyi{CRVAL}{#1}}
\newcommand{\CTYPE}[1]{\keyi{CTYPE}{#1}}
\newcommand{\CUNIT}[1]{\keyi{CUNIT}{#1}}

\newcommand{\keyii}[3]{\hbox{{\tt #1\hspace{1pt}{$#2$}\_{$#3$}}\/}}
\newcommand{\PC}[2]{\keyii{PC}{#1}{#2}}
\newcommand{\CD}[2]{\keyii{CD}{#1}{#2}}
\newcommand{\PV}[2]{\keyii{PV}{#1}{#2}}
\newcommand{\PS}[2]{\keyii{PS}{#1}{#2}}
\newcommand{\PCij}{\PC{i}{j}}
\newcommand{\CDij}{\CD{i}{j}}

\newcommand{\PVk}[1]{\hbox{{\tt PV\hspace{1pt}{$k$}\_{#1}{$a$}}\/}}

\newcommand{\keyv}[1]{\hbox{{\tt #1}}}

\begin{document}

\title{Representations of spectral coordinates in FITS}

\author{E. W. Greisen\inst{1} \and
        M. R. Calabretta\inst{2} \and
        F. G. Valdes\inst{3} \and
        S. L. Allen\inst{4}}

\institute{National Radio Astronomy Observatory,
           PO Box O,
           Socorro, NM 87801-0387, USA
\and       Australia Telescope National Facility,
           PO Box 76,
           Epping, NSW 1710, Australia
\and       National Optical Astronomy Observatories,
           PO Box 26732,
           Tucson, AZ 85719, USA
\and       UCO/Lick Observatory,
           University of California,
           Santa Cruz, CA 95064, USA}

\offprints{E. W. Greisen,\\\email{egreisen@nrao.edu}}

\date{{\bf Revised last 30 September 2005}, 05 May 2005,
    / Received  / Accepted }

\authorrunning{E. W. Greisen et al.}

\abstract{Greisen \&\ Calabretta (\cite{kn:GC}) describe a generalized
method for specifying the coordinates of FITS data samples.  Following
that general method, Calabretta \&\ Greisen (\cite{kn:CG}) describe
detailed conventions for defining celestial coordinates as they are
projected onto a two-dimensional plane.  The present paper extends the
discussion to the spectral coordinates of wavelength, frequency, and
velocity.  World coordinate functions are defined for spectral axes
sampled linearly in wavelength, frequency, or velocity, linearly in
the logarithm of wavelength or frequency, as projected by ideal
dispersing elements, and as specified by a lookup table.
}

\maketitle


\section{Introduction}

\label{s:intro}

The present paper is the third in a series of papers that
establishes methods by which information about the physical
coordinates of FITS data may be transferred along with the binary
image, random groups, and table data.  In ``Paper I'' Greisen \&\
Calabretta (\cite{kn:GC}) describe a revised method for transferring
coordinate information in the FITS header and outline some rules
governing the values assigned to the new standard header keywords.
In ``Paper II'' Calabretta \&\ Greisen (\cite{kn:CG}) specify the
conventions necessary to define celestial coordinates in a
two-dimensional projection of the sky.  This paper defines the
parameters and conventions needed to specify spectral information
including frequency, wavelength, and velocity.  In
addition to these basic conversions, a world coordinate function is
defined to describe ideal optical dispersers of several common
types.  Several concepts that were suggested by spectroscopic
issues are generalized to apply to all types of coordinate axes.  These
are a generalized description of non-linear algorithms, the {\tt -LOG}
and {\tt -TAB} non-linear algorithms, and an axis-naming keyword.  In
``Paper IV'', Calabretta et al.~(\cite{kn:CVGTDAW}) specify methods to
describe the distortions inherent in the image coordinate systems of
real astronomical data.

Paper I describes the computation of the world or physical coordinates
as a multi-step process.  The vector of pixel offsets from the
reference point is multiplied by a linear transformation matrix and 
then scaled to physical units.  Mathematically, this is given by
\begin{equation}
   x_i = s_i \,\, q_i  =  s_i \, \sum_{j=1}^N m_{ij} \,\, \left(
             p_j - r_j \right)  , \label{eq:pmijx}
\end{equation}
\noindent where $p_j$ are pixel coordinates, $r_j$ are pixel
coordinates of the reference point given by \CRPIX{j}, $m_{ij}$ is a
linear transformation matrix given either by \PCij\ or \CDij, $N$ is
the dimensionality of the WCS representation given by
\keyw{NAXIS}\ or \keyw{WCSAXES}, and $s_i$ is a scaling given
either by \CDELT{i}\ or by 1.0.

The final step in the computation is the conversion of these linear
relative coordinates into the actual physical coordinates.  The
conventions to be applied to spectral axes, i.e.~to frequency,
wavelength, and velocity axes, are described in this paper.


\section{Coordinate definition}
\label{s:basics}

\begin{figure*}
\begin{center}
\resizebox{\hsize}{!}{\includegraphics{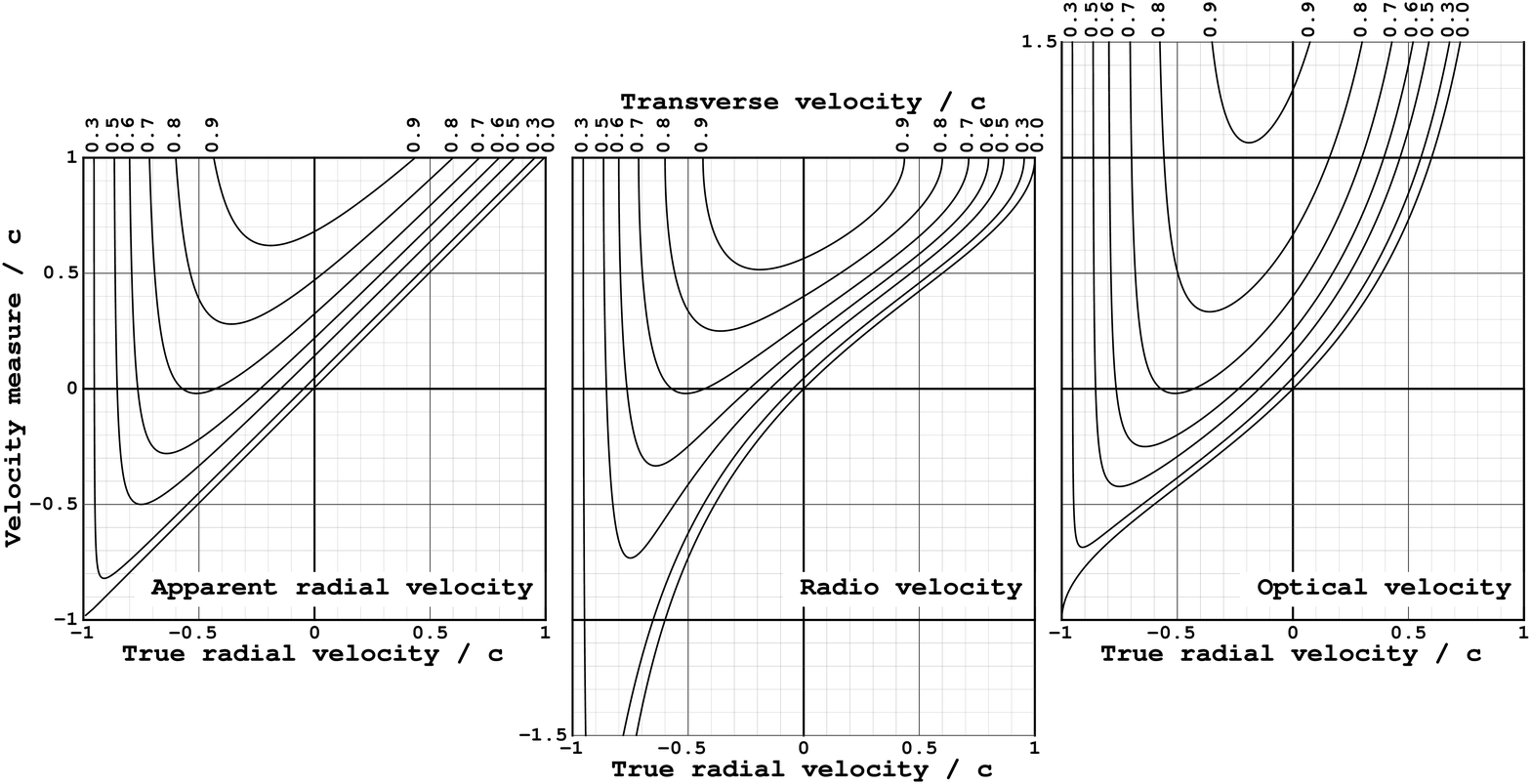}}
\caption[]{Conventional velocities as a function of true radial
velocity for selected values of the transverse velocity.  The apparent
radial velocity $\vv/c$ is plotted in the left panel, the radio
velocity $V/c$ is plotted in the center panel, and the optical
velocity $Z/c$ is plotted in the right panel. Note that, at a 
transverse velocity of $0.9 c$, the red shift $z$ always exceeds 1.
Note that each family of curves intersects the axis of zero
velocity measure at the same values of $\vv\sub{t}$ because the
redshift is zero for these combinations of $\vv\sub{t}$ and
$\vv\sub{r}$.  Thus, regardless of the value of $\vv\sub{t}$, for any
given redshift the velocity measures agree on whether the object {\em
appears} to be receding or approaching.  However, the object may
appear to be receding at high speed even when it is actually
approaching at high speed, though the converse is never true.  At
transverse velocities of $\vv\sub{t} / c > \sqrt{2} / 2 (\approx
0.71)$ all of the velocity measures are positive (receding) regardless
of whether the object is actually receding or approaching.
Furthermore, for $\vv\sub{r} / c$ below $-0.5$, the faster the object
approaches, the smaller the transverse velocity required to make it
appear to be receding!}
\label{fig:veltrans}
\end{center}
\end{figure*}

At this stage in the discussion, we consider the spectral world
coordinate at the reference point on all other axes.  This covers the
relatively simple case in which the spectral world coordinate is not
dependent on the other world coordinates.  Methods to describe
deviations from this assumption due to the choice of spectral
reference system are discussed in Sect.~\ref{s:reference}, while
instrumental distortions are discussed briefly in
Sect.~\ref{s:spatial} and, in more detail, in Paper~IV\@.

Spectral coordinates are commonly given in units of frequency,
wavelength, velocity, and other parameters proportional to these
three.  The coordinate types discussed here are then frequency,
wavelength, and ``apparent radial velocity'' denoted by the
symbols $\nu$, $\lambda$, and $\vv$.  There are also three
conventional velocities frequently used in astronomy.  These are
the so-called ``radio'' velocity, ``optical'' velocity, and redshift,
denoted here by $V, Z,$ and $z$ and given by $V = c (\nu_0 -
\nu)/\nu_0$, $Z = c (\lambda - \lambda_0)/\lambda_0$ and $z = Z / c$.
The velocities are defined so that an object receding from the
observer has a positive velocity.  We assume throughout that each
image has at most one spectral coordinate axis; a similar limit on
celestial coordinates was assumed in Paper II\@.

As discussed by Lindegren and Dravins (\cite{kn:LD}), the apparent
radial velocity to be described here is itself a conventional
velocity.  The shift of a spectral line from its rest frequency is
caused by a variety of effects, not just the Doppler shift due to a
radial motion.  The time dilation of an object moving with respect to
the observer causes a spectral shift, even if that motion is entirely
transverse.  Gravitational fields at the radiating object and along
the line of sight to the observer will shift the observed frequency.
Furthermore, the apparent spectroscopic velocity may be produced at
a peculiar point in the object, e.g.~upwelling convective cells,
rather than at the center of mass of the object.  It may also be
affected by absorption in intervening material and {in particular
by the cosmic redshift}.  We will use the term ``apparent radial
velocity'' here simply to refer to that pseudo-physical velocity
described by the equations presented here.  The apparent radial
velocity then is a sum of all spectroscopic effects presented as if
they were solely a radial velocity.  While this may be sufficiently
accurate for many uses, observers wishing to achieve very high
accuracies should consult Lindegren and Dravins, and references
therein, closely.

The effect of a transverse velocity deserves a little more discussion.
As given by Lang (\cite{kn:L}),
\begin{equation}
    \lambda = \lambda_0 \frac{c + \vv\sub{r}}{\sqrt{c^2 - \vv\sub{r}^2 -
               \vv\sub{t}^2}} ,\label{eq:Lang}
\end{equation}
where $\vv\sub{r}$ is the true radial velocity, $\vv\sub{t}$ is the
true transverse velocity, $c$ is the speed of light, and miscellaneous
effects such as gravitational redshifts are assumed negligible.  Even
at ordinary astronomical velocities, this has a measurable effect; a
transverse velocity of 100\,km\,s$^{-1}$ produces an apparent velocity
of 15\,m\,s$^{-1}$.  The effects of transverse velocities such as might
be found in astronomical jet sources on the apparent radial, radio,
and optical conventional velocities are illustrated in
Fig.~~\ref{fig:veltrans}.  When the transverse velocity is zero,
radio velocity ranges from $-\infty$ to $+c$, optical velocity from
$-c$ to $+\infty$, and apparent radial velocity is $\vv\sub{r}$ when
$\vv\sub{t} = 0$.  Because of all of the other processes that shift the
observed frequency, it is inappropriate to propose keywords to
describe true velocities at this time.  Additional discussion of
this matter is deferred to Appendix~\ref{appen:relativity}.

For the purposes of this section and the next, we consider those
spectral axes at a single celestial coordinate that are linearly
sampled in wavelength, frequency, or apparent radial velocity.
The radio and optical ``velocities'' are directly proportional to
frequency and wavelength, respectively, and thus do not constitute
additional cases for the type of sampling.  Frequency and wavelength
axes may also be regularly sampled in their logarithm.  Wavelengths
are sometimes given ``in air'' rather than in vacuum and denoted
here by $\lambda\sub{a}$.  This non-linear distinction is discussed in
Sect.~\ref{s:awav}.  Frequency may also be described in energy
($=h\nu$) units and in Kaysers (``wave number'' $=1/\lambda$) units.

Following the convention of Papers I and II, the first four
characters of \CTYPE{k\Ci} specify\footnote{While Papers I and II use
the generic intermediate axis number {\it i\/}, we use here the axis
number {\it k\/} as the spectral intermediate axis number.  The
single-character version code \Ci\ was introduced in Paper~I\@.  It
has values blank and {\tt A} through {\tt Z}\@.} the
coordinate type, the fifth character is {\tt '-'} and the next three
characters specify a predefined algorithm for computing the world
coordinates from intermediate physical coordinates.  When {\it k\/} is
the spectral axis, the first four characters shall be one of the codes
shown in Table~\ref{ta:codes}.  The mathematical symbols we will use
for the codes and their default units are also listed.  Units are
specified with the \CUNIT{k\Ci} keyword.  Standard values for unit
strings are defined in Paper~I\@.  The IAU-standard prefixes for
scaling the units are described in Paper I and should be used with all
coordinate types, except that the dimensionless ones are not scaled.

\begin{table}
\centering
   \caption{Spectral coordinate type codes, (characters 1--4 of
      \CTYPE{k\Ci}).}
   \renewcommand{\arraystretch}{1.25}
   \setlength\tabcolsep{5.4pt}
   \begin{tabular}{llccl}
      \hline\hline
      Code  & Name & Symbol & Associate  & Default \\
            &      &        & variable   & units   \\
      \hline
      \keyv{FREQ} & Frequency
                  & $\nu$
                  & $\nu$
                  & Hz \\
      \keyv{ENER} & Energy
                  & $E$
                  & $\nu$
                  & J  \\
      \keyv{WAVN} & Wavenumber
                  & $\kappa$
                  & $\nu$
                  & m$^{-1}$ \\
      \keyv{VRAD} & Radio velocity
                  & $V$
                  & $\nu$
                  & m\,\,s$^{-1}$ \\
      \keyv{WAVE} & Vacuum wavelength
                  & $\lambda$
                  & $\lambda$
                  & m \\
      \keyv{VOPT} & Optical velocity
                  & $Z$
                  & $\lambda$
                  & m\,\,s$^{-1}$ \\
      \keyv{ZOPT} & Redshift
                  & $z$
                  & $\lambda$
                  & -- \\
      \keyv{AWAV} & Air wavelength
                  & $\lambda\sub{a}$
                  & $\lambda\sub{a}$
                  & m \\
      \keyv{VELO} & Apparent radial velocity
                  & $\vv$
                  & $\vv$
                  & m\,\,s$^{-1}$ \\
      \keyv{BETA} & Beta factor ($\vv/c$)
                  & $\beta$
                  & $\vv$
                  & -- \\
      \hline
   \end{tabular}
   \label{ta:codes}
\end{table}

The non-linear algorithm in use is specified by the final 3
characters of \CTYPE{k\Ci}.  In spectral codes,\ the first of the
three characters specifies the physical parameter type in which the
data are regularly sampled, the second character is {\tt 2},\ and
the third character specifies the physical parameter type in which the
coordinate is expressed.  Non-linear algorithm codes, the last
four to be introduced later, are summarized in Table~\ref{ta:NLcodes}.
When the algorithm in use is linear, the last four characters of
\CTYPE{k\Ci}\ shall be blank. 
The original FITS paper by Wells et al.~(\cite{kn:WGH}) contained a
number of suggestions for the values of \CTYPE{i}.  Those suggestions
are to be superseded by the conventions of this paper and of
Papers~I and II\@.

\begin{table}
  \centering
   \caption{Non-linear algorithm codes, (characters 6--8 of
      \CTYPE{k\Ci}).}
   \renewcommand{\arraystretch}{1.25}
   \begin{tabular}{lll}
      \hline\hline
                 & Regularly      &                \\
      Code       & sampled in     & Expressed as   \\
      \hline
      \keyv{F2W} & Frequency      & Wavelength     \\
      \keyv{F2V} & Frequency      & Apparent radial velocity\\
      \keyv{F2A} & Frequency      & Air wavelength \\
      \keyv{W2F} & Wavelength     & Frequency      \\
      \keyv{W2V} & Wavelength     & Apparent radial velocity \\
      \keyv{W2A} & Wavelength     & Air wavelength \\
      \keyv{V2F} & Apparent radial velocity  & Frequency      \\
      \keyv{V2W} & Apparent radial velocity  & Wavelength     \\
      \keyv{V2A} & Apparent radial velocity  & Air wavelength \\
      \keyv{A2F} & Air wavelength & Frequency      \\
      \keyv{A2W} & Air wavelength & Wavelength     \\
      \keyv{A2V} & Air wavelength & Apparent radial velocity \\
      \hline
      \keyv{LOG} & logarithm &
                                Any 4-letter coordinate type \\
      \keyv{GRI} & detector &
                                Any from Table~\ref{ta:codes}\\
      \keyv{GRA} & detector &
                                Any from Table~\ref{ta:codes}\\
      \keyv{TAB} & not regular &
                                Any 4-letter coordinate type \\
      \hline
   \end{tabular}
   \label{ta:NLcodes}
\end{table}

The relationships between the basic physical quantities $\nu,
\lambda$, and $\vv$ are shown in Table~\ref{ta:speceqn}.  The
symbols $\lambda_0$ and $\nu_0$ are the rest wavelength and frequency,
respectively, of the spectral line used to associate velocity with
observed wavelength and frequency.  The relationships between the
derived quantities and the basic quantities with which they are
associated are shown in Table~\ref{ta:propeqn}.  Derivatives are
included in both tables since they will be needed in the coordinate
computation.


\section{Basic coordinate computation}
\label{s:define}

This section describes the computation of coordinates that are
sampled linearly or logarithmically in a coordinate of similar type,
and of coordinates that are sampled linearly in a coordinate of
different type.  The intermediate world coordinate for spectral axis
$k$, given by Eq.~(\ref{eq:pmijx}), will be denoted, for convenience,
by
\begin{equation}
   \vw \equiv x_k .
\label{eq:w}
\end{equation}
The final world coordinate will be denoted by $\vS$.  At the reference
point, it has the value $\vS\sub{r}$, which is defined by the
keyword \CRVAL{k\Ci}.


\subsection{Linear coordinates}
\label{s:linear}

Linear coordinates are represented in \CTYPE{k\Ci} by characters 1-4
containing any code in column one of Table~\ref{ta:codes} and
characters 5-8 blank.  The world coordinate value for spectral axis
{\it k} is computed with the above definitions and
Eq.~(\ref{eq:pmijx}) as
\begin{equation}
   \vS = \vS\sub{r} + \vw .
   \label{eq:linear}
\end{equation}

As a general rule, non-linear coordinate systems will be
constructed so that they satisfy Eq.~(\ref{eq:linear}) to first order
at the reference point.


\subsection{Logarithmic coordinates}
\label{s:LOG}

Data are often sampled in logarithmic increments in one or more
coordinates.  For example, spectra are sometimes sampled in
logarithmic increments of wavelength or frequency.  With this type of
sampling, the source motion, expressed as $V, Z,$ or $z$, shifts the
``pixel'' coordinates of spectral features by the same amount over the
whole image.  This allows relative velocities between spectra to be
determined using cross-correlation methods in the pixel arrays.

For logarithmic axes, the last four characters shall be {\tt
'-LOG'}\@.  While there are only three logarithmic\ coordinates
commonly used in spectroscopy, namely {\tt FREQ-LOG}, {\tt WAVE-LOG}
and {\tt AWAV-LOG}, it would be unwise to forbid any coordinate
type with the {\tt -LOG} non-linear algorithm.  Many such combinations
may have little physical meaning or be intractable mathematically, but
these are simply reasons to be cautious when using {\tt -LOG}\@.  This
algorithm is evaluated simply
\begin{equation}
  \vS  = \vS\sub{r}\, e^{\vw/\vS\sub{r}}  . \label{eq:log}
\end{equation}

This form of the logarithm satisfies the requirement that
$\Dfracr{\vS}{\vw} = 1$ so that Eq.~(\ref{eq:linear}) is satisfied to
first order at points near the reference point.  Thus $\vS \approx
\vS\sub{r} + \vw$.  This form of the logarithmic algorithm also has
the desirable attribute that the units of the coordinate ($\vS$),
reference coordinate, and scale are the same and are of a simple form.
The units for \CRVAL{k\Ci}, \CDELT{k\Ci}, and \CD{k}{j\Ci} are
specified by the \CUNIT{k\Ci} keyword as defined in Paper~I\@.  These
quantities are in normal, non-logarithmic units such as {\tt 'Hz'} for
{\tt FREQ-LOG} and {\tt 'm'} for {\tt WAVE-LOG}\@.  The prefixes and
alternate units described in Paper I may be used, such as {\tt 'MHz'}
and {\tt 'Angstrom'}.

Logarithmic quantities are frequently expressed as log base 10
rather than as natural logarithms.  For such data, the \CDELT{k\Ci}
and \CD{k}{j\Ci} will need to compensate by including a factor of
$\ln(10)$.

\begin{table}
\centering
\caption[]{Basic spectral transformation equations.}
\begin{tabular}{l@{\hspace{15pt}}l}
   \hline
   \noalign{\medskip}
   \Mc{\nu = \frac{c}{\lambda}
         \label{eq:B1}} &
      \Ma{\Dfrac{\nu}{\lambda} = -\frac{c}{\lambda^2}
         \label{eq:B2}} \\
   \Mc{\nu = \nu_0 \frac{c - \vv}{\sqrt{c^2 - \vv^2}}
         \label{eq:B3}} &
      \Ma{\Dfrac{\nu}{\vv} = -\frac{c\nu_0} {(c+\vv) \sqrt{c^2-\vv^2}}
         \label{eq:B4}} \\
   \Mc{\lambda = \frac{c}{\nu}
         \label{eq:B5}} &
      \Ma{\Dfrac{\lambda}{\nu} = -\frac{c}{\nu^2}
         \label{eq:B6}} \\
   \Mc{\lambda = \lambda_0 \frac{c + \vv}{\sqrt{c^2 - \vv^2}}
         \label{eq:B7}} &
      \Ma{\Dfrac{\lambda}{\vv} = \frac{c\lambda_0} {(c - \vv)
                                  \sqrt{c^2 - \vv^2}}
         \label{eq:B8}} \\
   \Mc{\vv = c \frac{\nu_0^2 - \nu^2}{\nu_0^2 + \nu^2}
         \label{eq:B9}} &
      \Ma{\Dfrac{\vv}{\nu} = -\frac{4c\nu\nu_0^2}{(\nu^2 + \nu_0^2)^2}
          \label{eq:B10}} \\
   \Mc{\vv = c \frac{\lambda^2-\lambda_0^2} {\lambda^2+\lambda_0^2}
          \label{eq:B11}} &
      \Ma{\Dfrac{\vv}{\lambda} = \frac{4c\lambda\lambda_0^2}
              {(\lambda^2 + \lambda_0^2)^2}
          \label{eq:B12}} \\
      \hline
   \end{tabular}
   \label{ta:speceqn}
\end{table}

\subsection{Coordinate axis names}
\label{s:CNAME}

The generality of this algorithm, and the {\tt -TAB} algorithm to
be introduced below, suggest the need for a more general description
of the coordinate than may be indicated in the first four letters in
the value of \CTYPE{k\Ci}\@.  We hereby reserve the keyword
\begin{center}
\begin{tabular}{l}
\noalign{\vspace{-5pt}}
\keyi{CNAME}{i\Ci} \hspace{2em} (character-valued)\\
\noalign{\vspace{-5pt}}
\end{tabular}
\end{center}
\noindent to provide a description of a particular coordinate in up to
68 characters.  Its default value will be all blank.  For binary table
vectors, the keyword will be {\it i\/}{\tt CNA}{\it n\Ci}, while for
pixel lists it will be {\tt TCNA}{\it n\Ci}, where $i$ is the
intermediate world coordinate axis number and $n$ is the table column
number to which the keyword applies.  This keyword may be used with
standard axis types such as {\tt GLON}/{\tt GLAT} or {\tt FREQ}, but
will be of greatest value with non-standard axis types such as,
hypothetically,\ \CTYPE{i} = {\tt 'TFPY-LOG'} meant to indicate
``log of Y position in telescope focal plane,'' which would be
recorded in the \keyi{CNAME}{i\Ci}\ card for that axis.  This keyword
provides a name for an axis in a particular WCS, while the {\tt
WCSNAME\Ci} keyword names the particular WCS as a whole.

\begin{table}
\centering
\caption[]{Extended spectral transformation equations.}
\begin{tabular}{l@{\hspace{15pt}}l}
   \hline
   \noalign{\medskip}
   \Mc{\nu = \nu_0 (1 - \frac{V}{c})
         \label{eq:P1}} &
      \Ma{\Dfrac{\nu}{V} = -\frac{\nu_0}{c}
         \label{eq:P2}} \\
   \Mc{V = c \frac{\nu_0 - \nu}{\nu_0}
          \label{eq:P3}} &
      \Ma{\Dfrac{V}{\nu} = -\frac{c}{\nu_0}
          \label{eq:P4}} \\
   \Mc{\nu = E / h
         \label{eq:P5}} &
      \Ma{\Dfrac{\nu}{E} = 1 / h
         \label{eq:P6}} \\
   \Mc{E = h \nu
          \label{eq:P7}} &
      \Ma{\Dfrac{E}{\nu} = h
          \label{eq:P8}} \\
   \Mc{\nu = c \kappa
         \label{eq:P9}} &
      \Ma{\Dfrac{\nu}{\kappa} = c
         \label{eq:P10}} \\
   \Mc{\kappa = \nu / c
          \label{eq:P11}} &
      \Ma{\Dfrac{\kappa}{\nu} = 1 / c
          \label{eq:P12}} \\
   \\
   \Mc{\lambda = \lambda_0 (1 + \frac{Z}{c})
         \label{eq:P13}} &
      \Ma{\Dfrac{\lambda}{Z} = \frac{\lambda_0}{c}
         \label{eq:P14}} \\
   \Mc{Z = c \frac{\lambda - \lambda_0}{\lambda_0}
          \label{eq:P15}} &
      \Ma{\Dfrac{Z}{\lambda} = \frac{c}{\lambda_0}
          \label{eq:P16}} \\
   \Mc{\lambda = \lambda_0 (1 + z)
         \label{eq:P17}} &
      \Ma{\Dfrac{\lambda}{z} = \lambda_0
         \label{eq:P18}} \\
   \Mc{z = \frac{\lambda - \lambda_0}{\lambda_0}
          \label{eq:P19}} &
      \Ma{\Dfrac{z}{\lambda} = \frac{1}{\lambda_0}
          \label{eq:P20}} \\
   \\
   \Mc{\vv = c \beta
         \label{eq:P21}} &
      \Ma{\Dfrac{\vv}{\beta} = c
         \label{eq:P22}} \\
   \Mc{\beta = \vv / c
          \label{eq:P23}} &
      \Ma{\Dfrac{\beta}{\vv} = 1 / c
          \label{eq:P24}} \\
   \hline
\end{tabular}
\label{ta:propeqn}
\end{table}


\subsection{Non-linear spectral coordinates}
\label{s:other}

We now consider the case where an axis is linearly sampled in spectral
variable $\vX$, but is to be expressed in terms of variable $\vS$.

\subsubsection{Spectral algorithm codes}

Given the large number of spectral variables in Table~\ref{ta:codes}
it is clear that there are very many pairwise combinations; in
general, the relationship between any pair may be non-linear.
However, each of the spectral variables in Table~\ref{ta:codes} is
linearly related to one or other of $\nu, \lambda$, $\lambda\sub{a}$,
or $\vv$.  Thus we may restrict $\vX$ to one of these four basic
variables.

Even with this restriction on $\vX$ there are still many possible
combinations of $\vX$ and $\vS$.  In order to reduce the number still
further we introduce an intermediate variable, denoted by $\vP$,
that is the basic variable, $\nu, \lambda$, $\lambda\sub{a}$, or
$\vv$, with which $\vS$ is associated via a linear relation.  This
associate variable is listed for each spectral variable in column 4 of
Table~\ref{ta:codes}.  Thus the sequence of transformations may be
summarized as 
$$p_j \rightarrow x_k\: (\equiv w) \rightarrow X \leadsto P \rightarrow S$$
where the non-linear transformation, indicated by the wiggly arrow,
is between $X$ and $P$.

Table~\ref{ta:speceqn} lists the equations, $\vX = \vX(\vP)$, and
their inverses, $\vP = \vP(\vX)$, linking the basic spectral types
$\nu, \lambda$, and $\vv$ (discussion of $\lambda\sub{a}$ is deferred
to Sect.~\ref{s:awav}).  These equations are generally non-linear.
Likewise, Table~\ref{ta:propeqn} lists the linear relations, $\vS =
\vS(\vP)$, and their inverses, $\vP = \vP(\vS)$, linking each spectral
variable with its associate. When $\vS$ is one of the basic types,
$\nu, \lambda$, $\lambda\sub{a}$, or $\vv$, then $\vP \equiv \vS$ and
$\vS(\vP)$ is identity.

Thus the functional relationship between $\vS$ and $\vX$ is specified
via intermediate variable $\vP$ as $\vS(\vX) = \vS(\vP(\vX))$ with
inverse $\vX(\vS) = \vX(\vP(\vS))$.  Since $\vS(\vP)$ is linear, $\vP$
and $\vX$ must always differ, otherwise $\vP(\vX)$ would be identity
with the result that $\vS(\vX)$ would be linear, contrary to the
assumption of a non-linear axis.

Non-linear spectral coordinate codes are constructed on this basis by
combining the spectral coordinate type code for $\vS$ from
Table~\ref{ta:codes} with the non-linear spectral algorithm code in
Table~\ref{ta:NLcodes}.  The first letter of the algorithm code
defines $\vX$ as frequency (\keyv{F}), wavelength (\keyv{W}), air
wavelength (\keyv{A}), or apparent radial velocity (\keyv{V}),
and the third letter likewise defines $\vP$.  For example, in
\keyv{ZOPT-F2W}, $\vX$ is \keyv{F}requency, $\vP$ is
\keyv{W}avelength, and the non-linear conversion between the two
(\keyv{F2W}) is defined in Table~\ref{ta:speceqn},
Eq.~(\ref{eq:B5}).  The desired spectral coordinate $\vS$ is redshift
(\keyv{ZOPT}), and this is related to associate variable $\vP$, the
\keyv{W}avelength, by the linear equation given in
Table~\ref{ta:propeqn}, Eq.~(\ref{eq:P19}).

It is apparent from the foregoing that it is possible to construct
invalid spectral {\tt CTYPE} coordinate codes.  For example,
\keyv{ZOPT-F2V} is unrecognized since $z$ is associated with
$\lambda$, not $\vv$.  That is not to say that this code could not be
interpreted in principle --- after all, equations linking $z$ and
$\vv$ could have been included in Table~\ref{ta:propeqn} ---
simply that it is not recognized in practice.  The associate variable,
$\vP$, was introduced in the first place to reduce the possible number
of combinations and adding new ones like this would defeat that
purpose.  This is particularly relevant to the software
implementation; \keyv{ZOPT-F2V} would tell software to chain its $\nu$
-- $\vv$ function with a $\vv$ -- $z$ function but, in general, the
latter function will not have been defined.

\subsubsection {Spectral algorithm chain}

Consider now the computation of spectral coordinate $\vS$ for an axis
that is linearly sampled in $\vX$.  The statement that an axis is
linearly sampled in $\vX$ simply means that
\begin{equation}
   \vX = \vX\sub{r} + \vw \, \Dfrac{\vX}{\vw}  ,
   \label{eq:X}
\end{equation}
where $\D{\vX}/\D{\vw}$ is constant.  This constant is determined by
imposing the requirement that
\begin{equation}
   \Dfracr{\vS}{\vw} = 1  ,
\end{equation}
so that Eq.~(\ref{eq:linear}) is satisfied to first order at points
near the reference point:
\begin{equation}
   \vS \approx \vS\sub{r} + \vw .
\end{equation}
Thus
\begin{equation}
   \Dfrac{\vX}{\vw} = \Dfracr{\vP}{\vS} / \Dfracr{\vP}{\vX} ,
   \label{eq:dXdw}
\end{equation}
which gives $\D{\vX}/\D{\vw}$ as a function of $\vX\sub{r}$,
$\D{\vP}/\D{\vS}$ being constant and $\D{\vP}/\D{\vX}$ a function
of $\vX$.  Given that the functions $\vS = \vS(\vP)$ and $\vP =
\vP(\vX)$ are known, as are their inverses $\vX = \vX(\vP)$ and $\vP =
\vP(\vS)$, then the equation for $\vS$ as a function of $\vw$ may be
obtained from Eqs.~(\ref{eq:X}) and (\ref{eq:dXdw})
\begin{equation}
   \vS(\vw) = \vS \left(
                     \vP \left(
                        \vX(\vP(\vS\sub{r})) +
                          \vw \Dfracr{\vP}{\vS} / \Dfracr{\vP}{\vX} \,
                     \right)
                  \right)  ,
   \label{eq:Xnlin}
\end{equation}
where $\vS\sub{r}$ is given by \CRVAL{k\Ci}.

Equation~(\ref{eq:Xnlin}) suggests a three-step algorithm chain:
\begin{enumerate}
\item Compute $\vX$ at $\vw$ using Eq.~(\ref{eq:X});
      $\vX\sub{r} = \vX(\vP(\vS\sub{r}))$ and $\Ddiv{\vX}{\vw}$ are
      constants that need be computed once only, the latter
      obtained from Eq.~(\ref{eq:dXdw}) as a function of
      $\vX\sub{r}$.
\item Compute $\vP$ from $\vX$ using the appropriate equation from
      Table~\ref{ta:speceqn}.
\item Compute $\vS$ from $\vP$ using the appropriate equation from
      Table~\ref{ta:propeqn}.
\end{enumerate}
The inverse computation by which the intermediate coordinate $\vw$ is
computed for a given value of $\vS$ is effected by traversing the
algorithm chain in the reverse direction.  The inverse equations
required are all listed in Tables~\ref{ta:speceqn} and
\ref{ta:propeqn}.  The inverse of Eq.~(\ref{eq:X}) is, trivially,
\begin{equation}
   \vw = (\vX - \vX\sub{r}) \, / \, \Dfrac{\vX}{\vw} \cdot
\end{equation}


\begin{table*}
   \centering
   \caption[]{Sample non-linear coordinate combinations, where
     $w$, the intermediate coordinate of Eq.~(\ref{eq:w}), has a
     different meaning for each of these equations  It has units given
     by the spectral coordinate type code as listed in
     Table~\ref{ta:codes}.}
   \begin{tabular}{llllll}
      \hline
      \noalign{\medskip}
      \TT{FREQ-W2F} &
      \Ma{\nu  = \left( \frac{\nu\sub{r}}{\nu\sub{r}-w} \right)
                       \nu\sub{r}} \\
      \TT{VELO-W2V} &
         \Ma{v = c\, \frac{A^2_\lambda-1}{A^2_\lambda+1}} & \Wh &
         \Md{A_\lambda \equiv \frac{c^2-v\sub{r}^2+cw}
                              {(c-v\sub{r})\sqrt{c^2-v\sub{r}^2}}} \\
      \TT{VRAD-W2F} &
         \Ma{V = \frac{V\sub{r} (c-V\sub{r}) + cw}{c-V\sub{r} + w}} \\
      \TT{WAVE-F2W} &
         \Ma{\lambda = \left( \frac{\lambda\sub{r}}
                        {\lambda\sub{r}-w} \right) \lambda\sub{r}} \\
      \TT{VELO-F2V} &
         \Ma{v = c\, \frac{A^2_\nu-1}{A^2_\nu+1}} & \Wh &
         \Md{A_\nu \equiv \frac{c^2-v\sub{r}^2-cw}{(c+v\sub{r})
                          \sqrt{c^2-v\sub{r}^2}}} \\
      \TT{VOPT-F2W} &
         \Ma{Z = \frac{Z\sub{r} (c+Z\sub{r}) + cw}{c+Z\sub{r}-w}
         \label{eq:Zw}} \\
      \TT{ZOPT-F2W} &
         \Ma{z = \frac{z\sub{r} (1+z\sub{r}) + w}{1+z\sub{r}-w}
         \label{eq:zw}} \\
      \TT{WAVE-V2W} &
         \Ma{\lambda = \rsl\sqrt{\frac{1+B_\lambda}{1-B_\lambda}}} &
                       \Wh &
         \Md{B_\lambda \equiv \frac{\lambda\sub{r}^4-\rsl^4 +
              4\rsl^2\lambda\sub{r}w}{(\rsl^2+\lambda\sub{r}^2)^2}} \\
      \TT{FREQ-V2F} &
         \Ma{\nu = \rsn\sqrt{\frac{1-B_\nu}{1+B_\nu}}} & \Wh &
         \Md{B_\nu \equiv \frac{\rsn^4-\nu\sub{r}^4-4\rsn^2\nu\sub{r}
                                      w} {(\rsn^2+\nu\sub{r}^2)^2}} \\
      \TT{VOPT-V2W} &
         \Ma{Z = c \left(\sqrt{\frac{1+D_Z}{1-D_Z}} - 1 \right)}
                 & \Wh &
         \Md{D_Z \equiv \frac{Y_Z^4 - 1 + 4Y_Zw / c}{(1+Y_Z^2)^2}}
                 & \Wi &
         \Mc{Y_Z \equiv 1 + Z\sub{r} / c} \\
      \hline
   \end{tabular}
   \label{ta:Nonlincc}
\end{table*}

\subsubsection{Example non-linear calculation}

As an example of a non-linear coordinate computation, consider a
\keyv{ZOPT-F2W} axis.  \keyv{F} in the \keyv{F2W} code indicates that
the axis is linearly sampled in frequency, but it is to be expressed
in terms of redshift as indicated by the spectral coordinate type of
\keyv{ZOPT}\@.  Table~\ref{ta:codes} indicates that redshift is
associated with wavelength, hence the \keyv{W} in \keyv{F2W} is
correct.  Of the $\vX$, $\vP$, and $\vS$ variables in the preceding
section, the axis is linear in frequency so $\vX$ is $\nu$, the
associated variable is wavelength, so $\vP$ is $\lambda$, and the
required variable is redshift, so $\vS$ is $z$.

Since \CRVAL{k\Ci} for the \keyv{ZOPT-F2W} axis would be recorded as a
redshift, $z\sub{r}$, this must first be converted to frequency by
applying Eqs.~(\ref{eq:P17}),
\begin{eqnarray*}
   \lambda\sub{r} & = & \lambda_0 (1 + z\sub{r})
\end{eqnarray*}
and (\ref{eq:B1}),
\begin{eqnarray*}
   \nu\sub{r} & = & \frac{c}{\lambda\sub{r}}
       \, = \, \frac{c}{\lambda_0 (1 + z\sub{r})} \cdot
\end{eqnarray*}
This provides $\vX\sub{r}$ for Eq.~(\ref{eq:X}).  Then $\Ddiv{\nu}{w}$
(i.e.\ $\Ddiv{\vX}{w}$) is obtained from Eqs.~(\ref{eq:P18}) \&
(\ref{eq:B6}) evaluated at the reference point:
\begin{eqnarray*}
   \Dfrac{\nu}{w} & = & \Dfracr{\lambda}{z} / \Dfracr{\lambda}{\nu}
        \, = \, \lambda_0 \, / \, \frac{-c}{\nu\sub{r}^2}
        \, = \, - \frac{\nu\sub{r}^2}{\nu_0}  \cdot
\end{eqnarray*}
Redshift may now be computed for any given value of $\vw$.  The first
step is to compute $\nu$ (i.e.\ $\vX$) at $\vw$ using
Eq.~(\ref{eq:X}):
\begin{eqnarray*}
   \nu & = & \nu\sub{r} + w \, \Dfrac{\nu}{w} \cdot
\end{eqnarray*}
In this instance $\vw$ will be a redshift.  Then $\lambda$ (i.e.\
$\vP$) may be computed from $\nu$ using Eq.~(\ref{eq:B5}) from
Table~\ref{ta:speceqn}:
\begin{eqnarray*}
   \lambda & = & c / \nu .
\end{eqnarray*}
The third and final step is to compute $z$ (i.e.\ $\vS$) from
$\lambda$ using Eq.~(\ref{eq:P19}) from Table~\ref{ta:propeqn}:
\begin{eqnarray*}
   z & = & \frac{\lambda - \lambda_0}{\lambda_0} \cdot
\end{eqnarray*}

It is also possible to combine the three steps into a single equation,
although there are many pair-wise combinations of spectral variables
and for some the combined equations may be quite cumbersome.  For the
sake of illustration, consider the above equations for the
\keyv{ZOPT-F2W} axis.  Eq.~(\ref{eq:X}) becomes
\begin{eqnarray*}
   \nu & = & \frac{c}{\lambda_0 (1 + z\sub{r})} -
             \frac{cw}{\lambda_0 (1 + z\sub{r})^2} \cdot
\end{eqnarray*}
Substituting this in the equations for $\lambda$ and $z$ and
simplifying we obtain
\begin{eqnarray*}
   z = \frac{z\sub{r} (1 + z\sub{r}) + w}{1 + z\sub{r} + w} \cdot
\end{eqnarray*}
A representative sample of these direct translation equations is shown
in Table~\ref{ta:Nonlincc}.  They are useful in showing the nature of
the coordinate non-linearity.

One thing to notice about the equations of Table~\ref{ta:Nonlincc} is
that the meaning of $\vw$ differs for each.  For example,
Eq.~(\ref{eq:Zw}) may not be obtained from Eq.~(\ref{eq:zw}) simply by
multiplying both sides by $c$: in Eq.~(\ref{eq:zw}) $\vw$ is a
redshift (\keyv{ZOPT}), whereas in Eq.~(\ref{eq:Zw}) it is an optical
velocity (\keyv{VOPT}).


\subsubsection{Coordinate parameters}
\label{s:COPAR}

Aside from \CRVAL{k\Ci}, the coordinate computations of
Sect.~\ref{s:other} require one extra parameter when evaluating the
\keyv{F2V}, \keyv{V2F}, \keyv{W2V}, \keyv{V2W}, \keyv{A2V}, and
\keyv{V2A} non-linear algorithms, namely the rest frequency or
wavelength of the spectral-feature of interest. These are fundamental
physical parameters so, rather than use the \PV{i}{ma} parameters
defined in Paper~I, which would tend to disguise them, the
special floating-valued keywords
\begin{center}
   \begin{tabular}{l}
      \noalign{\vspace{-5pt}}
      \keyw{RESTFRQ}\Ci\ \hspace{1em} (floating-valued), \\
      \noalign{\vspace{-5pt}}
   \end{tabular}
\end{center}
and
\begin{center}
   \begin{tabular}{l}
      \noalign{\vspace{-5pt}}
      \keyw{RESTWAV}\Ci\ \hspace{1em} (floating-valued), \\
      \noalign{\vspace{-5pt}}
   \end{tabular}
\end{center}
are hereby reserved for the purpose.  They are represented by
symbols $\nu_0$ and $\lambda_0$, respectively.  Their units are {\tt
'Hz'} and {\tt 'm'} respectively, fixed to save having additional
keywords to define them. \keyw{RESTWAV}\Ci\ is for wavelengths in
vacuum only. One or the other of these keywords must be included
for the above-mentioned algorithm codes; usually \keyw{RESTFRQ}\Ci\
would be given for \keyv{F2V} and \keyv{V2F}, and \keyw{RESTWAV}\Ci\
for the others.  FITS writers should always write one or other of
these when it is meaningful to do so, even for algorithm codes such as
\keyv{F2W} or \keyv{W2A} that do not require them.

Keyword \keyw{RESTFREQ} has been used in previous FITS files and
should be recognized as equivalent to \keyw{RESTFRQ}\@.


\section{Air wavelengths}
\label{s:awav}

The wavelengths so far discussed are measured in vacuum.  However,
dispersion coordinates for UV, optical, and IR spectra at $\lambda >
200$ nm are commonly given as wavelengths in air.  The relative
difference between the two varies between 0.028\%\ and 0.032\%\ across
this range.  To identify wavelengths measured in dry air at standard
temperature and pressure rather than in vacuo, we introduce the
coordinate type \keyv{AWAV} for which the reference value and
increment must be expressed accordingly.

The two measures of wavelength are related by
\begin{equation}
   \lambda =  n(\lambda\sub{a}) \lambda\sub{a}  , \\
   \label{eq:vacair}
\end{equation}
where $\lambda$ is the wavelength in vacuum, $\lambda\sub{a}$ is the
wavelength in air, and $n(\lambda\sub{a})$ is the index of refraction
of dry air at standard temperature and pressure.  This varies
non-linearly with wavelength.  The standard relation given by
Cox~(\cite{kn:Cox}) is mathematically intractable and somewhat dated.
The International Union of Geodesy and Geophysics~(\cite{kn:IUGG})
adopted the rather simpler formula\footnote{The quoted
formul\ae\ apply only at normal optical wavelengths.  In the
UV and IR spectral domains, atmospheric absorption lines cause
refractivity to be a strong, weather dependent function of frequency.
See, for example, Mathar~(\cite{kn:M}).}
\begin{equation}
 n(\lambda\sub{a}) = 1 + 10^{-6}\, \left(\, 287.6155 +
   \frac{1.62887}{\lambda\sub{a}^2} +
   \frac{0.01360}{\lambda\sub{a}^4}\, \right)
   \label{eq:RefIndx} ,
\end{equation}
where $\lambda\sub{a}$ is the wavelength in micrometers.  This formula
suffices for conversions from air to vacuum when no more than 0.25
parts per million accuracy is required.  The derivative, which may be
required in Eq.~(\ref{eq:dXdw}), is
\begin{equation}
   \Dfrac{\lambda}{\lambda\sub{a}} = 1 + 10^{-6}\, \left(\, 287.6155 -
       \frac{1.62887}{\lambda\sub{a}^2} -
       \frac{0.04080}{\lambda\sub{a}^4}\, \right)
   \label{eq:dlambda_a}   \cdot
\end{equation}

While the inversion of Eq.~(\ref{eq:RefIndx}) is algebraically
intractable, the vacuum wavelength may be used to evaluate
Eq.~(\ref{eq:RefIndx}) since it differs so little from the air
wavelength.  The resulting error in the index of refraction amounts to
$1:10^{9}$, and this is less than than the accuracy of the empirical
formula.  Thus,
\begin{equation}
   \lambda\sub{a} = \lambda / n(\lambda) .
   \label{eq:airvac}
\end{equation}

As usual, an axis that is linearly sampled and expressed in air
wavelengths is described with a \CTYPE{k\Ci}\ of \keyv{'AWAV'} and
evaluated with Eq.~(\ref{eq:linear}).

Algorithm codes for the non-linear conversions are \keyv{A2F},
\keyv{A2W}, \keyv{A2V}, and their complements, \keyv{F2A}, \keyv{W2A},
and \keyv{V2A} as listed in Table~\ref{ta:NLcodes}.  Use of the
three-step procedure described in Sect.~\ref{s:other} would require
$\lambda\sub{a}$ as a function of each of the other spectral
variables, together with their inverses and derivatives.  Thus it is
much simpler to handle air wavelengths as a separate, extra step in
the algorithm chain.  For example, to compute world coordinates for
\keyv{VRAD-A2V}, the value of \CRVAL{k\Ci} would first have to be
converted from radio velocity to air wavelengths via
Eqs.~(\ref{eq:P1}), (\ref{eq:B5}), and (\ref{eq:airvac}), and
$\Ddiv{\lambda\sub{a}}{\vw}$ for Eq.~(\ref{eq:dXdw}) would be obtained
from Eqs.~(\ref{eq:P2}), (\ref{eq:B2}), and (\ref{eq:dlambda_a}).
Then, the value of $\lambda\sub{a}$ computed for $\vw$ via
Eq.~(\ref{eq:X}) would be transformed to vacuum wavelengths via
Eq.~(\ref{eq:vacair}), and then to radio velocity via
Eqs.~(\ref{eq:B1}) and (\ref{eq:P3}).


\section{Dispersed spectra}

One common form of spectral data is produced by imaging the light from
a {\em disperser}.  The wavelengths of the light at some position in
the image are related to the position and wavelength of the light in
the field of view illuminating the disperser; the relation is defined
by the physics of the disperser.  In general the light received at a
pixel in the detector is a superposition of different wavelengths from
different points in the field of view.  However, if the field of view
is limited spatially, usually by aperture masks and fiber optics, each
pixel receives light from only a small range of wavelengths.  It is
then possible to define a world coordinate function relating pixel
position and effective wavelength.  This is the basis of many
astronomical spectrographs.

In the following section we use the physical relation applicable to
the dispersers commonly used in astronomical spectrographs to define a
world coordinate function for computing spectral coordinates.  The
relation applies to the simple, though common, case of single
dispersers.  More complex spectrographs with multiple dispersers, such
as those using multiple passes through prisms, are not described
by the methods of this section.  The equations developed below
also assume that the radiation enters perpendicular to the face of the
prism, a condition not met by some widely used spectrometers.
Alternatives to using this ideal world coordinate function, based on
the physics of simple dispersers, are the table lookup described
in Sect.~\ref{s:tlookup2} and empirical function fits provided by
the the distortion function mechanism described in Paper~IV\@.

We require that the dispersion occurs along just one direction on the
detector and that the intermediate coordinates are computed so that
only one world coordinate axis corresponds to wavelength.  Paper IV
describes how distortions and effects of the aperture shape can be
removed to satisfy this requirement.  The distortion correction is
also used to remove aberrations causing departures from the ideal
physical behavior of the dispersers assumed here.

\begin{figure*}
   \centerline{\includegraphics[height=138pt]{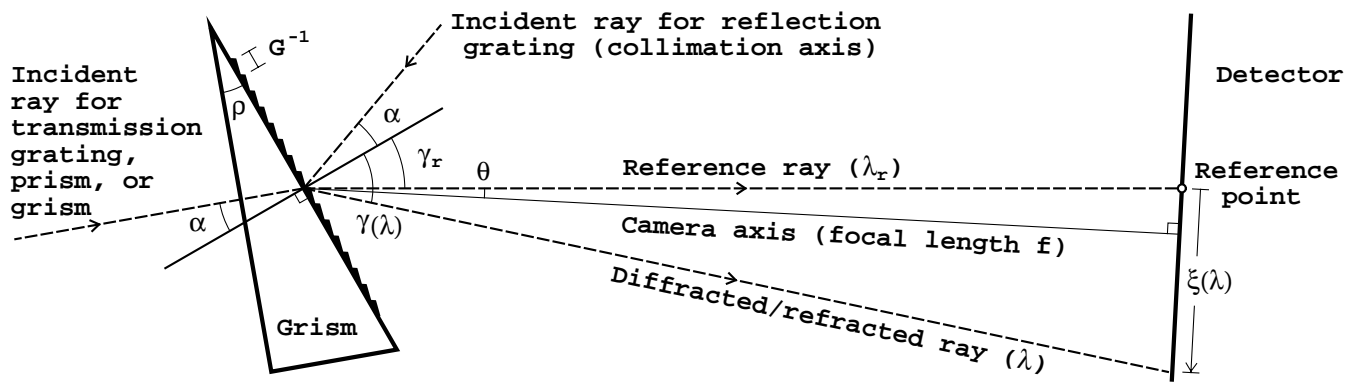}}
   \caption{Geometry of gratings, prisms, and grisms.  This simplified
       representation omits the collimation and focusing optics.
       Dashed lines mark ray paths in the plane of the figure -- the
       ``dispersion plane''.  The normal to the grating/exit prism
       face and the normal to the detector plane are each projected
       onto the dispersion plane, and angles $\alpha$, $\gamma$, and
       $\theta$ are measured with respect to these projected normals.
       Usually the incident ray for a prism or grism is perpendicular
       to the entry face so that $\alpha$ is equal to the prism angle,
       $\rho$.  Angle $\gamma$ is wavelength-dependent, and
       consequently so is the offset $\xi$ in the dispersion direction
       on the detector. The intermediate spectral world coordinate,
       $\vw$, is proportional to $\xi$.  Reference wavelength
       $\lambda\sub{r}$ follows the reference ray defined by
       $\gamma\sub{r}$ and illuminates the reference point at $\vw =
       \xi = 0$.  The normal to the detector plane is shown tilted by
       angle $\theta$ from the reference ray though typically this
       angle is zero.  The grating spacing $G^{-1}$ is indicated.}
   \label{fig:grism}
\end{figure*}

The dispersers we consider are {\em gratings}, {\em prisms}, and {\em
grisms}, which are a combination of a prism with a grating within or
on the surface of the prism.  Gratings may be reflecting or
transmissive and may be fabricated with surface relief rulings,
holographic surface relief patterns, and holographic volume phase
patterns.  In the discussion we use the terms {\em grooves}, {\em
lines}, and {\em ruling} to refer to the periodic diffraction
structures that produce the interference.  By combining the two
physical equations for interference and refraction a single world
coordinate function covers all these cases with appropriate choice of
parameters.


\subsection{The grism coordinate function}
\label{sec:griscoord}

This section defines a world coordinate function using the basic laws
of refraction and interference.  It is beyond the scope of this
discussion to give the full background for the laws and concepts
underlying dispersive spectroscopy.  Of the many texts on the subject,
a standard reference is {\it Astronomical Optics} by
Schroeder~(\cite{kn:Sch}).

Spectroscopic dispersers are based on interference and refraction
effects which are naturally described in terms of wavelength.
Moreover, they are most often used in the regime where wavelength is
commonly the spectral coordinate.  For these reasons the grism
function is defined in terms of wavelength.  The function assumes that
all of the dispersion occurs at one place.  This is why
combinations of dispersers, other than a single grism, are not
described by this function.  This assumption means that for prisms and
grisms the function is rigorously correct only when the light enters
perpendicular to the face of the prism and the grating is at the exit
face.  However, even when these conditions are not exactly met the
function may still be a good approximation with slight adjustments of
the parameters.


\subsubsection{The grism equation}
\label{sec:griseqn}

The basic physical relationship between the wavelength, $\lambda$, the
angle of incidence of the light, $\alpha$, and the
diffracted/refracted angle, $\gamma$, is given by a combination
of the grating interference equation and Snell's law of refraction:
\footnote{Also known as Descartes' Law of Sines; see for example,
{\tt http://wikipedia.org/wiki/Snells\_law}}
\begin{equation}
    \frac{G m \lambda}{\cos\epsilon} = n(\lambda) \sin\alpha +
          \sin\gamma  ,     \label{eq:grism}
\end{equation}
where $G$ is the grating ruling density, $m$ is the interference
order, and $n(\lambda)$ is the wavelength-dependent index of
refraction of the prism material.  For a pure prism the order $m$ is
zero and for a reflection or transmission grating $n(\lambda)$ is the
unit function.  The plane containing $\alpha$ and $\gamma$ is the {\em
dispersion plane}, illustrated in Fig.~\ref{fig:grism}, and the angles
are measured relative to the projection in the dispersion plane of the
normal to the grating or exit prism face.  Usually the normal does lie
in the dispersion plane, but small values of $\epsilon$, the angle
between the normal and the dispersion plane, are sometimes used for
instrumental design reasons.

By convention, the angles of incidence and diffraction/refraction are
measured {\em from} the normal {\em to} the beam and $\alpha$ is
always measured positive in the anticlockwise direction.  The sign
changes on either side of the normal and this determines the sign of
$\gamma$.  Thus, for the reflection grating in Fig.~\ref{fig:grism},
$\alpha > 0$ and $\gamma < 0$, and vice versa for a transmission
grating, prism, or grism.

Reflection grating geometry is sometimes defined by the angle $\phi$
measured from the camera axis to the collimator axis and the tilt $t$
of the bisector relative to the grating normal.  If $\phi$ and $t$
obey the same sign convention as $\alpha$, then $\alpha = \phi / 2 +
t$.

In spectrographs with prisms or grisms the requirement that the
incident light be normal to the prism entry face means that $\alpha$
is equal to the prism angle ($\rho$ in Fig.~\ref{fig:grism}).  Even
for oblique entry, identifying $\alpha$ with the prism angle is often
a good initial approximation.

The requirement that the diffraction and refraction occur at one point
means that $\alpha$ is fixed and independent of wavelength.  The
variation of $\gamma$ with wavelength then defines the relation between
wavelength and position $\xi$ on the detector.  The reference
wavelength $\lambda\sub{r}$ is the wavelength at the reference point
corresponding to the zero point of $\xi$ and the intermediate spectral
world coordinate $\vw$.

The prism's dispersive power derives from the variation of its index
of refraction with wavelength.  While this variation depends on the
material and is generally non-linear, it is sufficient to approximate
it by a first-order Taylor expansion about the reference wavelength,
$\lambda\sub{r}$:
\begin{equation}
    n(\lambda) \approx n\sub{r} + n'\sub{r} \, (\lambda -
               \lambda\sub{r})  ,     \label{index}
\end{equation}
where we have written $n\sub{r}$ as a shorthand for
$n(\lambda\sub{r})$, and likewise $n'\sub{r}$ for
$\Ddivr{n}{\lambda}$.  Combining Eqs.~(\ref{eq:grism}) and
(\ref{index}) yields the grism equation
\begin{equation}
   \lambda = \frac{(n\sub{r} - n'\sub{r} \lambda\sub{r}) \sin\alpha +
                    \sin\gamma}
                  {Gm/\cos\epsilon - n'\sub{r} \sin\alpha}  ,
   \label{eq:lambdagamma}
\end{equation}
where the denominator must not be zero, though $Gm$, $n'\sub{r}$, and
$\alpha$ may be zero in different types of spectrographs.

In order to define a world coordinate function we need $\lambda$ as a
function of the intermediate world coordinate, $\vw$, which is
proportional to $\xi$.  Since Eq.~(\ref{eq:lambdagamma}) gives
$\lambda$ as a function of $\gamma$ it remains to determine the
relationship between $\gamma$ and $\xi$.  First note that
Eq.~(\ref{eq:lambdagamma}) evaluated at $\lambda = \lambda\sub{r}$
provides
\begin{equation}
   \gamma\sub{r} = \sin^{-1} \left( G m \lambda\sub{r} / \cos\epsilon -
                             n\sub{r} \sin\alpha \right)  ,
   \label{eq:gamma_r}
\end{equation}
the exit angle of the reference ray for which $\vw = \xi = 0$, as in
Fig.~\ref{fig:grism}.

Fig.~\ref{fig:grism} shows angle $\theta$, which is measured
from the reference ray to the camera axis using the same sign
convention as $\gamma$, i.e.~if $\gamma$ is clockwise-positive as for a
grism then so is $\theta$. Normally $\theta$ would be zero, but it is
included here to provide a small correction.  Depending on the sign
convention for $\gamma$, simple geometry for a flat detector gives
\begin{equation}
    \gamma = \gamma\sub{r} + \theta + \tan^{-1} (\pm \xi / f -
                \tan\theta)  ,       \label{eq:gamma}
\end{equation}
where $f$ is the effective focal length of the camera.  The plus sign
is taken when the sign convention for $\gamma$ is such that $\xi$
increases as $\gamma$ increases, and the minus sign otherwise.
Section~\ref{sec:griscale} discusses the resolution of this potential
sign ambiguity.


\subsubsection{\keyw{GRI} coordinate axes}
\label{sec:grisaxes}

In keeping with the preceding sections we wish to define general
grating, prism, and grism world coordinate representations such as
\keyv{WAVE-GRI}, \keyv{FREQ-GRI}, etc.

Bearing in mind that the grism equation, Eq.~(\ref{eq:lambdagamma}), is
expressed in terms of wavelength, then given a \keyv{FREQ-GRI} axis,
for example, it would be straightforward to convert the reference
frequency, $\nu\sub{r}$, given by \CRVAL{k}, from frequency to
wavelength via $\lambda = c / \nu$.  However, it would not be valid to
convert the intermediate spectral world coordinate, $\vw$, from
frequency to wavelength like this because it is not a true frequency.
While it may be a close approximation near the reference point it
deviates at points away from it.

Thus, interpretation of an axis such as \keyv{FREQ-GRI} necessarily
involves a procedure similar to that of Sect.~\ref{s:other}, and to
apply that methodology we need a parameter, the {\em grism parameter},
$\Gamma$, that is a known function of the spectral variables and for
which the axis is linearly sampled; this will substitute for $\vX$ in
Eq.~(\ref{eq:X}).  Since $\xi$ is proportional to $\vw$
\begin{equation}
   \xi = \sigma \vw  ,
\end{equation}
where $\sigma$ is a constant.  Combining this with Eq.~(\ref{eq:gamma})
we have
\begin{equation}
   \tan(\gamma - \gamma\sub{r} - \theta) = -\tan\theta \pm \vw \sigma /
              f .      \label{eq:grisparm}
\end{equation}
Thus the grism parameter may be identified as
\begin{equation}
   \Gamma = \tan(\gamma - \gamma\sub{r} - \theta)  ,
   \label{eq:Gammagamma}
\end{equation}
and this satisfies Eq.~(\ref{eq:X}) at the reference point:
\begin{equation}
   \Gamma\sub{r} = -\tan\theta.
   \label{eq:Gamma_r}
\end{equation}
The grism parameter has a simple geometrical interpretation in
Fig.~\ref{fig:grism}; it is the offset on the detector from the point
of intersection of the camera axis measured in units of the effective
focal length, $f$.

Following Sect.~\ref{s:other} (with $\vX$ replaced by $\Gamma$), we
have
\begin{equation}
   \Gamma = \Gamma\sub{r} + w \Dfrac{\Gamma}{\vw}
   \label{eq:Gammaw}
\end{equation}
where $d\Gamma/d\vw$ is constant:
\begin{equation}
   \Dfrac{\Gamma}{\vw} = \Dfracr{\Gamma}{\vP}
                        \Dfracr{\vP}{\vS}
                        \Dfracr{\vS}{\vw} \cdot
   \label{eq:dGdw}
\end{equation}
As before, $\vS$ is the spectral variable in which the axis, and hence
$\vw$, is expressed, and $\vP$ is the basic variable, $\nu, \lambda$,
$\lambda\sub{a}$, or $\vv$, with which $\vS$ is most closely
associated.  Recognized spectral variables and their associates are
listed in Table~\ref{ta:codes}.  When $\vS$ is one of the basic types,
$\nu, \lambda$, $\lambda\sub{a}$, or $\vv$, as is often the case, then
$\vP \equiv \vS$ whence $\Ddivr{\vP}{\vS} = 1$.

As previously, we require that \CDELT{i} or \CDij\ be set so that
\begin{equation}
   \Dfracr{\vS}{\vw} = 1
\end{equation}
(exactly how this is done is addressed in Sect.~\ref{sec:griscale}).
Since we only have $\Gamma$ directly in terms of $\gamma$ we may use
\begin{equation}
   \Dfracr{\Gamma}{\vP} = \Dfracr{\Gamma}{\lambda}
                          \Dfracr{\lambda}{\vP}  ,
\end{equation}
where $\Ddivr{\Gamma}{\lambda}$ itself may be deduced from the
derivatives of Eqs.~(\ref{eq:Gammagamma}) \& (\ref{eq:lambdagamma}):
\begin{eqnarray}
   \Dfracr{\Gamma}{\lambda}
      & = & \Dfracr{\Gamma}{\gamma} / \Dfracr{\lambda}{\gamma}
                   \nonumber  \\
      & = & \frac{Gm / \cos\epsilon - n'\sub{r} \sin\alpha}
                 {\cos\gamma\sub{r} \cos^2\theta} \cdot
   \label{eq:dGdlambda}
\end{eqnarray}
Combining Eqs.~(\ref{eq:dGdw}) to (\ref{eq:dGdlambda}) we have
\begin{equation}
   \Dfrac{\Gamma}{\vw} = \frac{Gm / \cos\epsilon - n'\sub{r}
               \sin\alpha}    {\cos\gamma\sub{r} \cos^2\theta}
               \Dfracr{\lambda}{\vP}     \Dfracr{\vP}{\vS}  ,
   \label{eq:dGammadw}
\end{equation}
where $\Ddivr{\lambda}{\vP}$ and $\Ddivr{\vP}{\vS}$ come from
Tables~\ref{ta:speceqn} \& \ref{ta:propeqn} respectively.  In the
common case where $\vS$, and hence $\vP$, is wavelength then
$\Ddivr{\lambda}{\vP}$ and $\Ddivr{\vP}{\vS}$ are both unity.

The foregoing provides everything needed to compute $\Gamma$ from
$\vw$, and this forms the first step of a five-step algorithm chain
for computing $\vS$ from $\vw$:
\begin{enumerate}
\item Compute $\Gamma$ at $\vw$ using Eq.~(\ref{eq:Gammaw}).
      $\Gamma\sub{r}$ and $\Ddiv{\Gamma}{\vw}$ are constants that
      need be computed once only; $\Gamma\sub{r}$ from
      Eq.~(\ref{eq:Gamma_r}) and $\Ddiv{\Gamma}{\vw}$ from
      Eq.~(\ref{eq:dGammadw}) using the appropriate equations from
      Tables~\ref{ta:speceqn} \& \ref{ta:propeqn}.
\item Compute $\gamma$ from $\Gamma$ using the inverse of
      Eq.~(\ref{eq:Gammagamma}):
      \begin{equation}
         \gamma = \tan^{-1}(\Gamma) + \gamma\sub{r} + \theta .
      \end{equation}
\item Compute the wavelength $\lambda$ from $\gamma$ via
      Eq.~(\ref{eq:lambdagamma}).
\item Compute $\vP$ from $\lambda$ using the appropriate equation from
      Table~\ref{ta:speceqn}.
\item Compute $\vS$ from $\vP$ using the appropriate equation from
      Table~\ref{ta:propeqn}.
\end{enumerate}
If $\vS$, and hence $\vP$, is wavelength then the last two steps are
not performed.

As usual, the inverse computation is effected by traversing the
algorithm chain in the reverse direction.  The inverse equations
required for the backward steps have already been given except that
for Eq.~(\ref{eq:lambdagamma}):
\begin{eqnarray}
   \gamma & = & \lefteqn{\sin^{-1}
            (\lambda (G m / \cos\epsilon - n'\sub{r} \sin\alpha) \  -}
               \nonumber \\*
      &   & \hskip 35pt (n_r - n'\sub{r} \lambda\sub{r}) \sin\alpha) ,
\end{eqnarray}
and also the inverse of Eq.~(\ref{eq:Gammaw}) which is trivial.

Grism parameters required for these calculations are provided by the
\PV{k}{ma} keywords defined in Table~\ref{ta:griparms}, with
$\gamma\sub{r}$ given by Eq.~(\ref{eq:gamma_r}), and $\lambda\sub{r}$
computed from the coordinate reference value, \CRVAL{k}, by the
appropriate equations from Tables~\ref{ta:speceqn} \&
\ref{ta:propeqn}.  Default values for missing parameters are also
defined in the table.  Generally only the first three parameters will
appear for gratings and the first five for grisms.

\begin{table}
\centering
   \renewcommand{\arraystretch}{1.25}
   \caption{Grism parameters, their default values, and required
     units.}
   \begin{tabular}{lcc}
      \hline\hline
      Keyword & Default & Units \\
      \hline
      \PVk{0} = $G$         & 0 & m$^{-1}$ \\
      \PVk{1} = $m$         & 0 &          \\
      \PVk{2} = $\alpha$    & 0 & deg  \\
      \PVk{3} = $n\sub{r}$  & 1 &          \\
      \PVk{4} = $n'\sub{r}$ & 0 & m$^{-1}$ \\
      \PVk{5} = $\epsilon$  & 0 & deg  \\
      \PVk{6} = $\theta$    & 0 & deg  \\
      \hline
   \end{tabular}
   \label{ta:griparms}
\end{table}


\subsubsection{Determination of the grism scale}
\label{sec:griscale}

A question raised above was how \CDELT{i} or \CDij\ may be set by a
WCS composer so that $\Ddivr{\vS}{\vw} = 1$.

From Eqs.~(\ref{eq:pmijx}) \& (\ref{eq:w})
\begin{equation}
   \vw = s_k q_k
\end{equation}
so
\begin{eqnarray}
   s_k & = & \Dfrac{\vw}{q_k} \, = \,
             \Dfracr{\vw}{\vS}
             \Dfracr{\vS}{\vP}
             \Dfracr{\vP}{\lambda}
             \Dfracr{\lambda}{q_k}.
   \label{eq:griscale}
\end{eqnarray}
Now, $\Ddivr{\vw}{\vS} = 1$ by design, $\Ddivr{\vS}{\vP}$ and
$\Ddivr{\vP}{\lambda}$ come from Tables~\ref{ta:propeqn} \&
\ref{ta:speceqn} respectively, and $\Ddivr{\lambda}{q_k}$ is the
wavelength dispersion (e.g.\ in nm/pixel) measured at the reference
point and this is an instrumental parameter.  Thus we have everything
needed to compute $s_k$.  When $\vS$ is wavelength, then all but the
last factor in Eq.~(\ref{eq:griscale}) is unity, and $s_k =
\Ddivr{\lambda}{q_k}$.

It is the correct choice of sign for $\Ddivr{\lambda}{q_k}$ that
resolves the sign ambiguity in Eq.~(\ref{eq:grisparm}).


\subsubsection{\keyv{GRA}: grisms in air}
\label{sec:grisair}

Thus far we have ignored the distinction between vacuum and air
wavelengths in the discussion of grism world coordinates.  In fact,
the $\lambda$ variable that appears in the grism equation may be
either, and in general $n(\lambda)$ in Eq.~(\ref{eq:grism}) is the
index of refraction of the prism material with respect to the
surrounding medium, either air or vacuum.

If the dispersion takes place in vacuum, as it does, for example, in a
spectrograph on a spacecraft, then the grism equation is correct for
vacuum wavelengths; if in air, it is correct for air wavelengths and
$\lambda$ in Sects.~\ref{sec:griseqn}, \ref{sec:grisaxes}, and
\ref{sec:griscale} should be replaced everywhere by $\lambda\sub{a}$.
Then as discussed in Sect.~\ref{s:awav}, the computation of quantities
such as $\Ddivr{\lambda\sub{a}}{\vP}$ in Eq.~(\ref{eq:dGammadw}) is
best handled by using the vacuum wavelength, $\lambda$, as an
intermediary, effectively introducing an extra step into the algorithm
chain.

In order to distinguish between grisms operated in air and in vacuum
we hereby reserve the use of \keyv{GRI} exclusively for vacuum
operation and introduce \keyv{GRA} for operation in dry air at
standard temperature and pressure.

Note that for real spectrographs operated in air at an observatory,
the actual wavelength system that \keyv{GRA} describes is for air at
the local conditions.  The coordinate value and increment at the
reference point are normally adjusted to the values at standard
conditions during calibration.  If the accuracy of the wavelength
measurement requires it, any further correction between wavelength at
local conditions and wavelength at standard conditions may be
accomplished via the methods of Paper~IV\@.


\subsection{\keyv{AWAV-GRA} examples}

\begin{figure}
\centerline{\includegraphics[width=8cm]{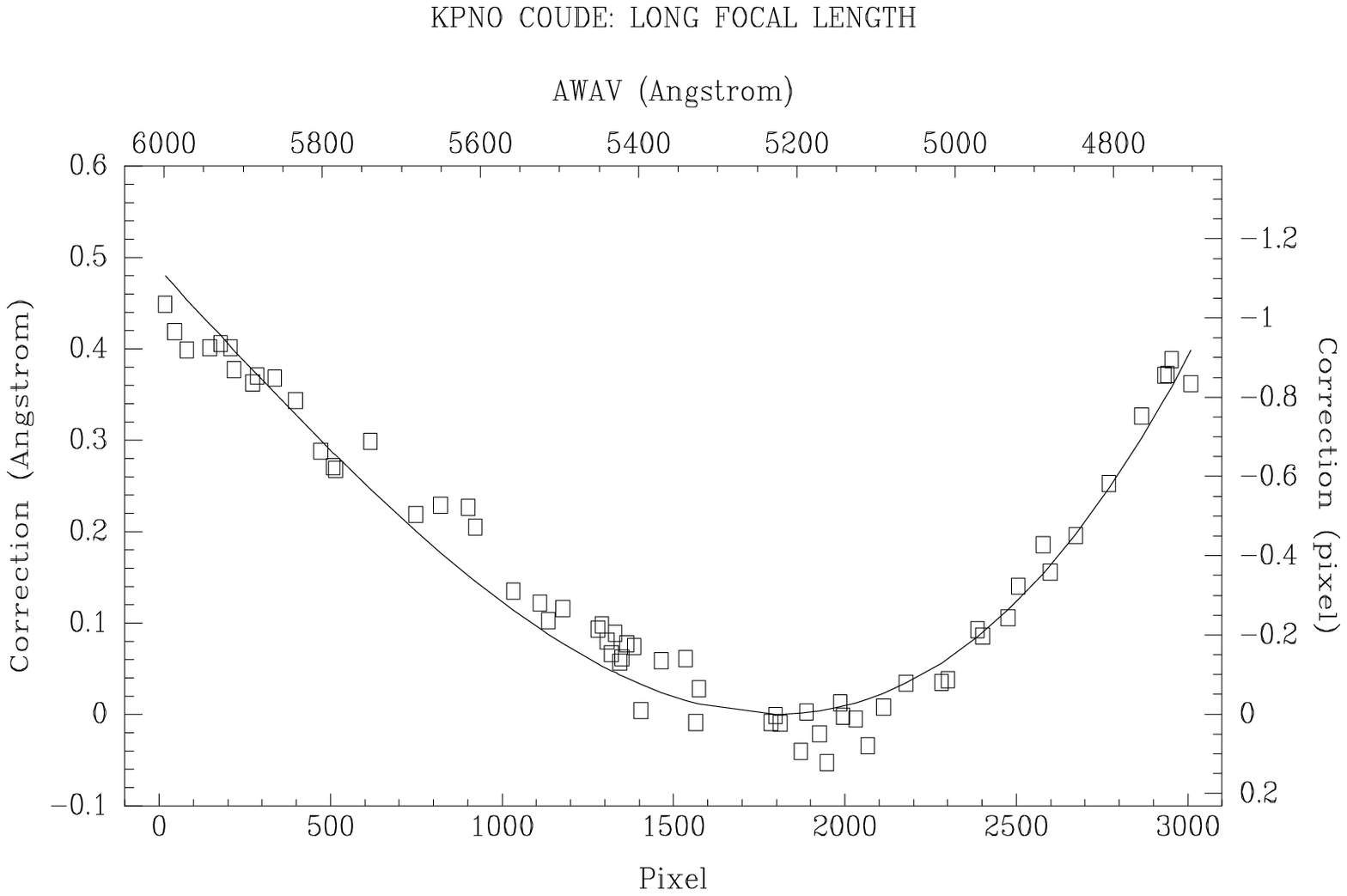}}
\verb!CTYPE1  = 'AWAV-GRA' / Grating dispersion function! \\
\verb!CUNIT1  = 'Angstrom' / Dispersion units! \\
\verb!CRPIX1  =     1801.7 / [pixel] Reference pixel! \\
\verb!CRVAL1  =     5225.2 / [Angstrom] Reference value ! \\
\verb!CDELT1  =    -0.4334 / [Angstrom/pixel] Dispersion! \\
\verb!PV1_0   =     3.16E5 / [m^(-1)] Grating density! \\
\verb!PV1_1   =          1 / Diffraction order! \\
\verb!PV1_2   =       13.9 / [deg] Incident angle! \\
\caption{KPNO Coud\'{e} Feed spectrograph with a long focal length
    and a 3K CCD\@.}
\label{Coude}
\end{figure}

\begin{figure}
\centerline{\includegraphics[width=8cm]{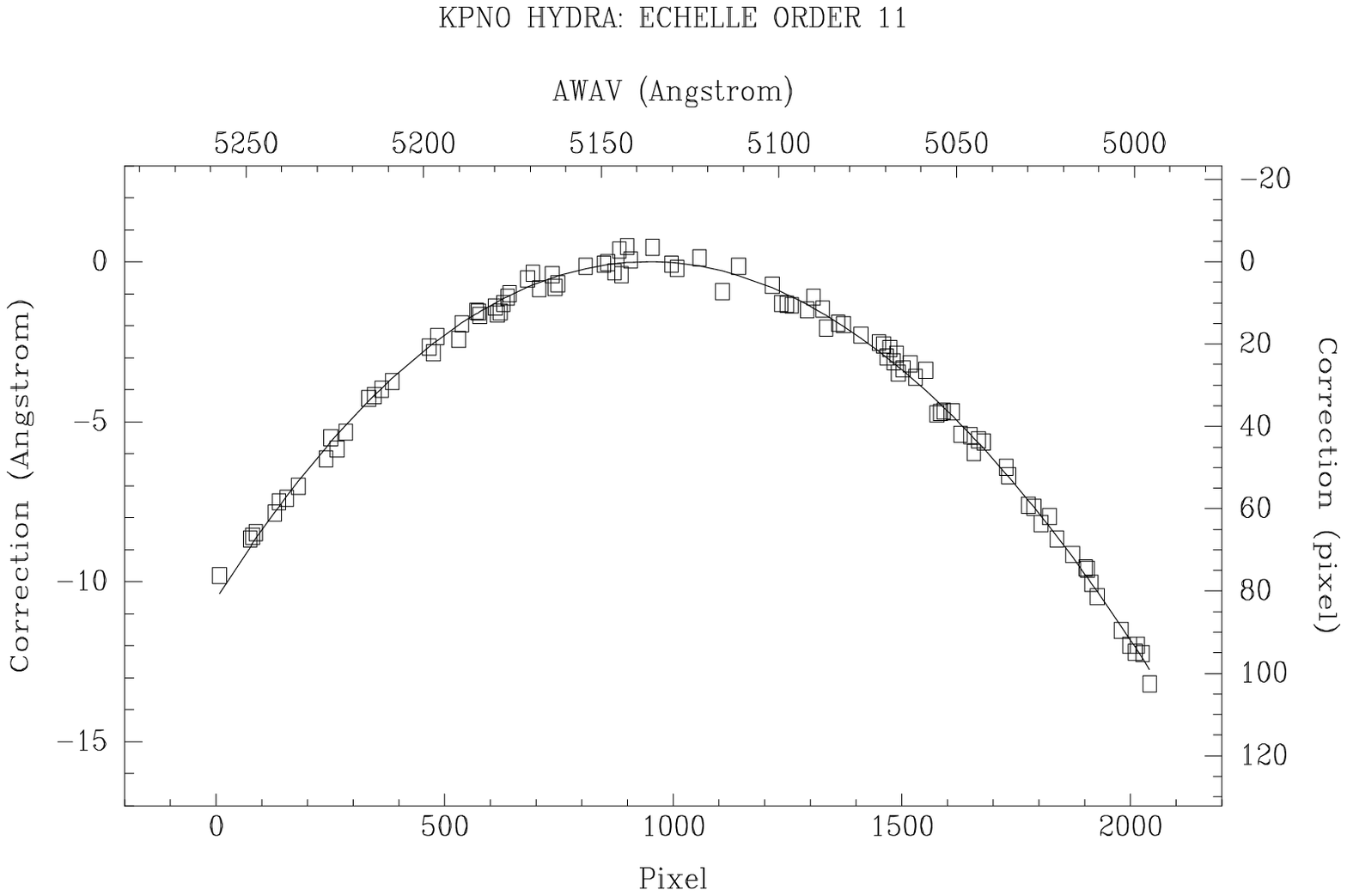}}
\verb!CTYPE1  = 'AWAV-GRA' / Grating dispersion function! \\
\verb!CUNIT1  = 'Angstrom' / Dispersion units! \\
\verb!CRPIX1  =      944.8 / [pixel] Reference pixel! \\
\verb!CRVAL1  =     5136.8 / [Angstrom] Reference value ! \\
\verb!CDELT1  =    -0.1287 / [Angstrom/pixel] Dispersion! \\
\verb!PV1_0   =     3.16E5 / [m^(-1)] Grating density! \\
\verb!PV1_1   =         11 / Diffraction order! \\
\verb!PV1_2   =       64.8 / [deg] Incident angle! \\
\caption{KPNO Hydra Fiber Bench Spectrograph using an echelle grating
    in $11^{\rm th}$ order with a 2K CCD\@.}
\label{Hydra}
\end{figure}

\begin{figure}
\centerline{\includegraphics[width=8cm]{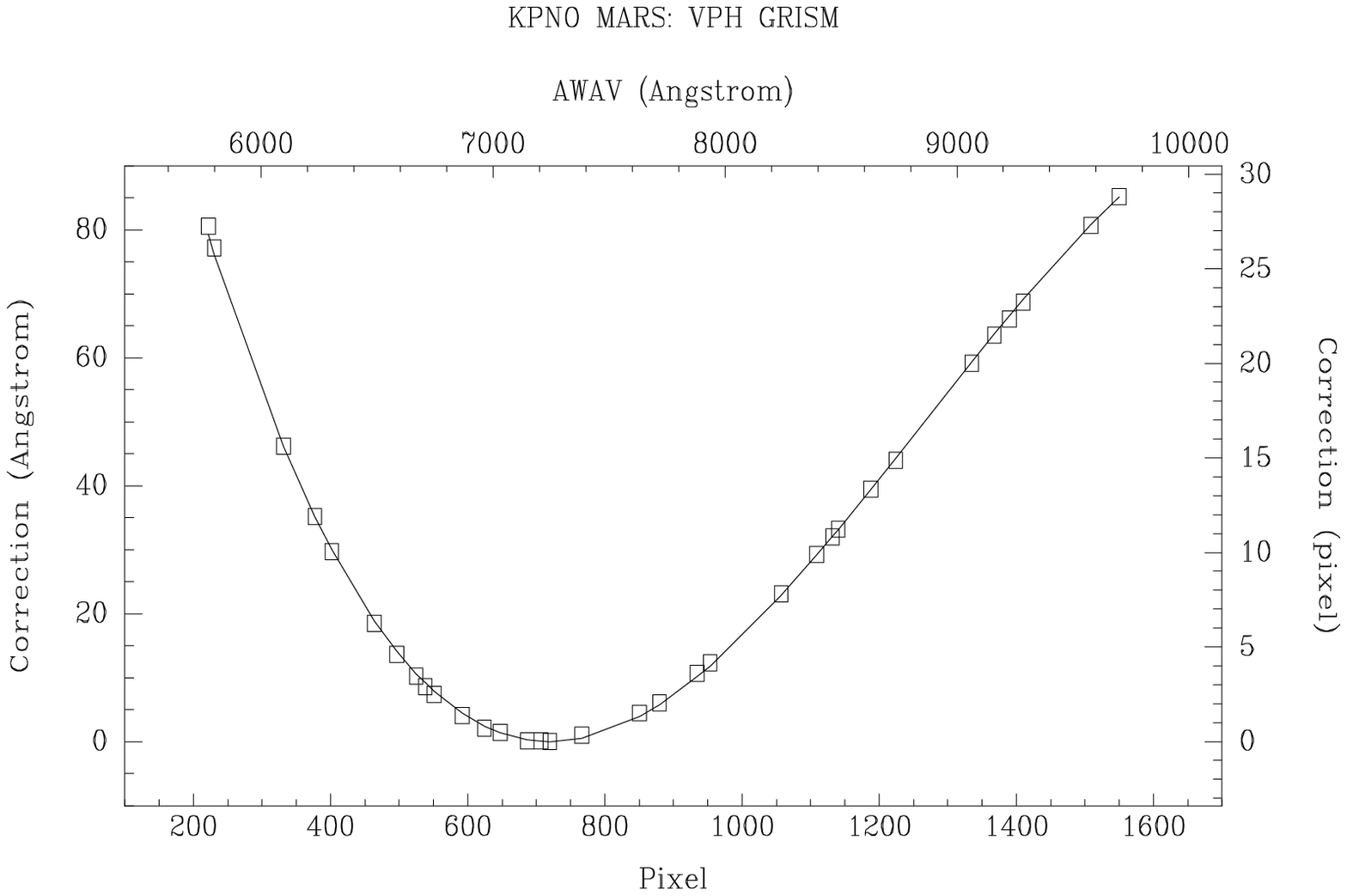}}
\verb!CTYPE1  = 'AWAV-GRA' / Grating dispersion function! \\
\verb!CUNIT1  = 'Angstrom' / Dispersion units! \\
\verb!CRPIX1  =      719.8 / [pixel] Reference pixel! \\
\verb!CRVAL1  =     7245.2 / [Angstrom] Reference value! \\
\verb!CDELT1  =      2.956 / [Angstrom/pixel] Dispersion! \\
\verb!PV1_0   =     4.50E5 / [m^(-1)] Grating density! \\
\verb!PV1_1   =          1 / Diffraction order! \\
\verb!PV1_2   =       27.0 / [deg] Incident angle! \\
\verb!PV1_3   =      1.765 / Reference refraction! \\
\verb!PV1_4   =   -1.077E6 / [m^(-1)] Refraction deriv! \\
\caption{KPNO MARS spectrograph with a 450 lines/mm volume phase
holographic grism and a 2K CCD\@.}
\label{MARS}
\end{figure}

This section illustrates application of the \keyv{GRA} world
coordinate function with three real-world examples of spectral images
from Kitt Peak National Observatory (KPNO) spectrographs
(Figs.~\ref{Coude}--\ref{MARS})\@.  Each spectrum is of an arc
calibration lamp that produces emission lines of known
wavelength.  The position of each line in the image along the spectral
world coordinate axis is measured by centroiding on the spectral line
profile and identified with the known rest wavelength to create a list
of pixel positions and wavelengths.

Each example figure shows plots of the positions and wavelengths for a
particular spectrograph.  The pixel positions are plotted along the
bottom axis.  Rather than plot the wavelengths directly, where it
would be difficult to see the shape of the curve, the difference or
correction between the known (air) wavelengths and the simple linear
world coordinate function {\tt AWAV} ($\lambda\sub{a}=\vS\sub{r}+\vw$)
are shown.  The corrections are plotted along the left axes.  In these
examples the wavelength units are consistently Angstroms.

The top axis is labeled with simple {\tt AWAV} linear
coordinates.  The right axis divides the wavelength correction by
\keyw{CDELT1}, the linear dispersion at the reference pixel.  This
represents the shift, in pixels, of the wavelengths on the detector
relative to where they would occur in a spectrograph with a linear
dispersion relation.

The departure of the data points from zero indicate the magnitude of
the world coordinate errors that would occur by using the simple
linear \keyv{AWAV} world coordinate function.  The solid lines in
the figures are the difference between the wavelengths produced by the
\keyv{AWAV-GRA} grating coordinate function and the linear \keyv{AWAV}
coordinate function evaluated at the pixel positions in the image.

The usefulness of the ideal grating coordinate function is that the
curves go through the measured data points with an appropriate choice
of parameters.  The parameters are essentially those known for the
spectrograph and disperser with small adjustments to some of the
parameters to produce the best fit to the calibration data.
(These small adjustments may be viewed as corrections for the
simplifications in the optical model.)  In these examples the
dispersion function represented by the grating world coordinate
relation is as good as typically provided by empirical polynomial
functions.  Higher order effects due to aberrations are corrected by
the distortion correction methods defined in Paper~IV\@.

A close reading of the equations above will reveal that the seven
grism parameters listed in Table~\ref{ta:griparms} are not
independent.  We have chosen these parameters because of their
physical meaning.  However, the independent parameters are $G m /
\cos(\epsilon), n\sub{r} \sin(\alpha), n'\sub{r} \sin(\alpha),$ and
$\theta$.  It is these combinations of parameters which must be used
in fitting data.

Each figure shows the arc line measurements, the coordinate function
curve, and the relevant world coordinate keywords used to produce the
curve.  Not all of the WCS keywords are shown.

The first example shows the behavior of a long focal length
spectrograph with a reflection grating used at a low angle of
incidence.  Because of the long focal length the deviation from
linearity is relatively small though still clearly significant.  The
grating has a density of 316 lines/mm and is used in first order
to produce a spectrum centered near 5225.2\,\AA\ with a dispersion of
0.43\,\AA/pixel.

The next example uses a 316 lines/mm echelle reflection grating, a
grating designed for use at large angles from the grating normal,
operated in a higher order.  It is used to produce a spectrum centered
near 5136.8\,\AA\ with a dispersion of 0.13\,\AA/pixel.  The departure
from linearity is in the opposite sense from the other examples
because the echelle grating is used with the diffracted angle at a
greater angle than the incidence angle.  The higher order results in a
fairly large departure from a linear WCS\@.

The last example illustrates use of a grism.  However, this is not the
more common grism with a ruled grating on the output face and the
input face oriented normal to the incident beam.  Instead, the grating
is sandwiched in the middle of the prism.  The prism is oriented with
the beam entering and leaving the prism at equal angles to the faces
resulting in a straight through configuration as with a standard
grism.

Another unusual feature of this grism is that it uses a volume phase
holographic (VPH) grating.  While the intensity response is different
from a surface relief grating (ruled or holographic) the dispersion
behavior is the same.

The full optical equation is complex even in this symmetric
configuration, but as shown in the figure using the \keyv{GRA}
function, a very good description of the coordinates can be obtained.

The prism has a $27\degr$ angle with an index of refraction of 1.764
near the reference wavelength.  The grating has an equivalent
interference pattern of 450 lines/mm.  Using these values and
adjusting the derivative of the index of refraction produces
Fig.~\ref{MARS}.


\section{Coordinates by table lookup}
\label{s:tlookup2}

There are numerous instances in which a physical coordinate is well
defined at each pixel along an image axis, but the relationship of the
coordinate values between pixels cannot be described by a simple
functional form.  An obvious example of this would be a
three-dimensional image consisting of a sequence of
two-dimensional images of an astronomical object recorded at an
arbitrary sequence of times determined in part by weather and time
assignment committees.  As another example, the calibration of some
spectrographs, such as those employing a diode array detector, is
represented best by a list of frequencies or wavelengths for each
pixel on the spectral axis rather than some functional form.


\subsection{{\tt {\it xxxx\/}-TAB} non-linear algorithm}

To support such representations for primary images and image
extensions, we define a table-lookup form for the value of
\CTYPE{i\Ci} as {\tt {\it xxxx\/}-TAB}, where {\it xxxx\/} are four
letters representing the type of coordinate, e.g.\ {\tt TIME} or {\tt
FREQ}\@.  As in Paper I, which established the ``4--3'' convention
for \keyi{CTYPE}{i}, the coordinate {\it xxxx\/} is a ``real''
coordinate, such as {\tt FREQ}, not an intermediate coordinate, such
as {\tt FREQ-F2W}, requiring an additional linear or non-linear
algorithm in order to be converted into a physical coordinate.

\subsubsection{{\tt -TAB} indexing concepts}

Consider first the case of a single (one-dimensional) coordinate
axis.  The {\tt -TAB} algorithm uses a list of coordinate values, the
{\em coordinate array}, to record coordinate values of the
appropriate type for the coordinate axis.  A second list of the same
length, the {\em indexing vector}, may be used in addressing the
coordinate array.  The indexing vector provides one level of
indirection which may be used to vary the sampling frequency of the
coordinate array along the coordinate axis.  The coordinate could
then be sampled at smaller intervals over that portion of the range in
which the instrument is more non-linear and sampled more coarsely over
regions in which it is better behaved.  The indexing algorithm, based
on linear interpolation, is defined in Sect.~\ref{s:tabinterp}.  If the
indexing vector is absent, it is taken to have values
$1,2,3,\ldots,K$, where $K$ is the number of values in the coordinate 
array.  See Sect.~\ref{s:TABradio} for an example of the use of the
indexing vector.

The concept described above covers separable (one-dimensional) axes
only.  It may be extended simply to $M$ non-separable axes so long as
the indexing vectors for each of the $M$ axes are separable.  $M$
coordinate values are required for each of the possible index
positions.  Therefore, the coordinates will be in a single 
$(1+M)$-dimensional array.  This {\em coordinate array} will have
dimensions $(M, K_1, K_2,\ldots K_M)$, where $K_m$ is the maximum
value of the index on axis $m+1$ of the coordinate array.  For
simplicity, degenerate axes are forbidden; therefore, $K_m > 1$.
The indexing vectors for each of the $M$ axes each contain a
one-dimensional array of length $K_m$.

The data in the indexing vectors must be monotonically increasing or
decreasing, although two adjacent index values in the vector may have
the same value.  However, it is not valid for an index value
to appear more than twice in an index vector, nor for an index value
to be repeated at the start or at the end of an index vector.
Furthermore, repeated index values are allowed only in the case of
one-dimensional separable axes.  (See the following section for a
discussion of the reasons for these limitations and how interpolation
is done in such cases; see also Sect.~\ref{s:Steve} for an example of
equal index values and further discussion of interpolation.)
Application programs must not sort the index and coordinate arrays
since this makes the relative order of the two equal index values
indeterminate.  The requirement for monotonic index values should
eliminate any need for sorting.  Note that it does not imply any
ordering of the values in the coordinate array.

\begin{figure}
  \centerline{
    \resizebox{8cm}{!}{ \Large
      \begin{pspicture}(400pt,417pt)
        \psset{xunit=1pt,yunit=1pt,runit=1pt}
        \psset{linewidth=2.5}
        \psframe(0,228)(400,418)
        \psframe(0,0)(400,220)
        \psset{linewidth=1.4}
        \rput(200,402){{\LARGE {\tt -TAB} without an indexing vector}}
        \rput(200,363){\psframebox{$\psi_m = \keyi{CRVAL}{i}\, +\,
            \sum\limits_{j=1}^{N}
            \keyii{CD}{i}{j}\, (p_j - \keyi{CRPIX}{j})$}}
        \qline(200,341)(200,326)
        \rput(200,316){single-row table \keyii{PS}{i}{0}}
        \qline(200,308)(200,298)
        \rput(200,288){$\psi_m$ is a direct index into table cell
              coordinate array}
        \psline(200,279)(200,269)(180,269)(180,259)
        \psline(176,263)(180,259)(184,263)
        \rput(242,266){\keyii{PS}{i}{1}}
        \rput(200,248){\psframebox{$C_1\, C_2\, C_3\, C_4\, C_5\,
              C_6\, C_7\, \ldots\, \, C_K$}}
        \rput(200,204){{\LARGE {\tt -TAB} with an indexing vector}}
        \rput(200,165){\psframebox{$\psi_m = \keyi{CRVAL}{i}\, +\,
           \keyi{CDELT}{i} \,\,   \sum\limits_{j=1}^{N}
           \keyii{PC}{i}{j}\, (p_j - \keyi{CRPIX}{j})$}}
        \qline(200,143)(200,128)
        \rput(200,120){single-row table \keyii{PS}{i}{0}}
        \qline(200,113)(200,103)
        \rput(200,95){Find index $\Upsilon_m$ by interpolating
            $\psi_m$ in table cell indexing vector $\Psi_k$}
        \psline(200,87)(200,78)(113,78)(113,68)
        \psline(109,72)(113,68)(117,72)
        \rput(60,76){\keyii{PS}{i}{2}}
        \rput(340,76){\keyii{PS}{i}{1}}
        \rput(108,57){\psframebox{$\Psi_1\, \Psi_2\, \Psi_3\,
            \Psi_4\, \Psi_5\, \Psi_6\, \Psi_7\, \ldots\, \, \Psi_K$}}
        \rput(292,57){\psframebox{$C_1\, C_2\, C_3\, C_4\, C_5\,
            C_6\, C_7\, \ldots\, \, C_K$}}
        \psline(113,48)(113,22)(128,22)
        \psline(124,26)(128,22)(124,18)
        \rput(207,29){$\Upsilon_m$ selects point in table}
        \rput(207,15){cell coordinate array $C_{\Upsilon_m}$}
        \psline(284,22)(299,22)(299,46)
        \psline(295,42)(299,46)(303,42)
      \end{pspicture}
      }
   }
\caption[]{{\tt -TAB} logic flow with and without an indexing vector.
The coordinate array subscript $m$ associated with intermediate
world coordinate axis $i$ is specified with keyword \PV{i}{{\tt 3}}.
In the case of an independent {\tt -TAB} axis it would have value 1.}
Note that $\psi_m$ is computed either with the \keyii{CD}{i}{j} 
form of the linear transformation matrix or the \keyii{PC}{i}{j} plus
\keyi{CDELT}{i} form.
\label{fig:TABasic}
\end{figure}
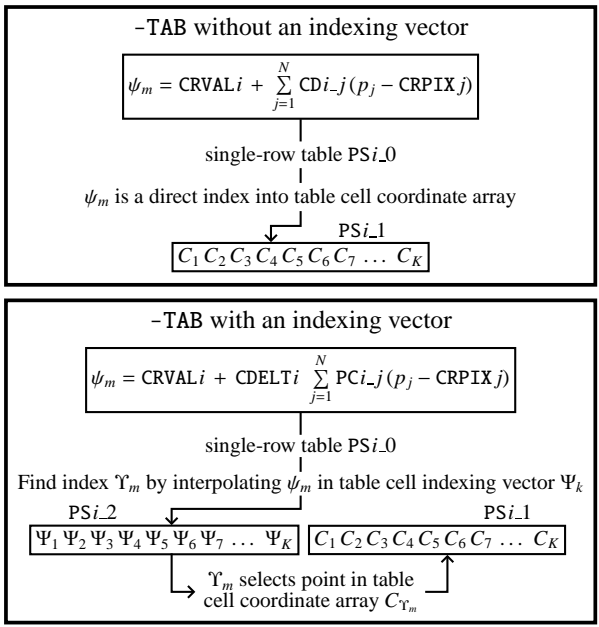

\subsubsection{Computing {\tt -TAB} coordinate values}
\label{s:tabinterp}

The indexing algorithm is illustrated schematically in
Fig.~\ref{fig:TABasic} for the one-dimensional case.  In the general
case, to determine the $M$ non-separable coordinate values $C_m$, one
first determines the $M$ indices, $\psi_m$ for addressing the
appropriate location in the table.  If intermediate world coordinate
axis $i$ is associated with the  $m^{\rm th}$ axis in the coordinate
array, one evaluates
\begin{equation}
     \psi_m = x_i + \mbox{\CRVAL{{i}\Ci}} ,
\end{equation}
where $x_i$ is computed following the prescriptions of
Eq.~(\ref{eq:pmijx}).   Using linear interpolation, if necessary, in
the indexing vector for intermediate world coordinate axis $i$, one
determines the location, $\Upsilon_m$, corresponding to $\psi_m$.
Then the coordinate value, $C_m$, of type specified by the first four
characters of \CTYPE{{i}\Ci}, is that at location $(m, \Upsilon_1,
\Upsilon_2,\ldots \Upsilon_M)$ in the coordinate array, again using
linear interpolation as needed.

In detail, the algorithm for computing $\Upsilon_m$, and thence
$C_m$, is as follows.  Scan the indexing vector, $(\Psi_1, \Psi_2,
\dots)$, sequentially starting from the first element, $\Psi_1$, until
a successive pair of index values is found that encompass $\psi_m$
(i.e. such that $\Psi_k \leq \psi_m \leq \Psi_{k+1}$ for monotonically
increasing index values or $\Psi_k \ge \psi_m \ge \Psi_{k+1}$ for 
monotonically decreasing index values for some $k$).  Then, when 
$\Psi_{k} \neq \Psi_{k+1}$, interpolate linearly on the indices
\begin{equation}
   \Upsilon_m = k + \frac{\psi_m - \Psi_k}{\Psi_{k+1} -
       \Psi_k}  ,\label{eq:indxcoord}
\end{equation}
which yields an index into the coordinate array.  However, if
$\Psi_k = \Psi_{k+1} (= \psi_m)$ then the result is undefined.

In the case where an index value is equal to $\psi_m$, the algorithm
above will find the interval $\Psi_k < \psi_m = \Psi_{k+1}$ for
monotonically increasing index values or $\Psi_k > \psi_m =
\Psi_{k+1}$ for monotonically decreasing index values, except when
$\psi_m = \Psi_1$.  Since two consecutive index values may be equal,
the index following the matched index must be examined.  If
$\Psi_{k+2} = \Psi_{k+1} = \psi_m$ (or $\Psi_2 = \Psi_1 = \psi_m$),
then $\Upsilon_m$ and $C_m$ are undefined. 

Linear interpolation via Eq.~(\ref{eq:indxcoord}) applies for $\psi_m$
in the range $\Psi_1$ to $\Psi_K$ inclusive.  Outside this range, for
$K > 1$, linear extrapolation is allowed for values of $\psi_m$ such
that $\Psi_1 - (\Psi_2 - \Psi_1) / 2 \le \psi_m < \Psi_1$ or
$\Psi_K < \psi_m \le \Psi_K + (\Psi_K - \Psi_{K-1}) / 2$ for monotonic
increasing index values, and for
$\Psi_1 + (\Psi_1 - \Psi_2) / 2 \ge \psi_m > \Psi_1$ or
$\Psi_K > \psi_m \ge \Psi_K - (\Psi_{K-1} - \Psi_K) / 2$ for monotonic
decreasing index values.  Extrapolation is also allowed for $K = 1$
with $\psi_m$ in the range $\Psi_1 - 0.5 \le \psi_m \le \Psi_1 + 0.5$
(noting that $\Psi_1$ should be equal to 1 in this case) whence
$\Upsilon_m = \psi_m$.

The value of $\Upsilon_m$ derived from $\psi_m$ must lie in the range
$0.5 \le \Upsilon_m \le K + 0.5$.  These extrapolation limits permit
assignment of coordinates to the regions of the pixels on the boundary
of the array which are outside of the centers of the boundary pixels
but within the conceptual ``edge'' of the boundary pixels.  In the
case of a single separable coordinate with $1 \le k \le \Upsilon_m <
k+1 \le K$, the coordinate value is given by
\begin{equation}
   C_m = C_k \, + \, \left( \Upsilon_m - k \right)\,
           \left( C_{k+1} - C_k \right)  .\label{eq:tabcoord}
\end{equation}
For $\Upsilon_m$ such that $0.5 \le \Upsilon_m < 1$ or
$K < \Upsilon_m \le K + 0.5$ linear extrapolation is permitted, with
$C_m = C_1$ when $K = 1$.

Conceptually, to compute the change in coordinate value between
$\Psi_k$ and $\Psi_{k+1}$, in the case when $\Psi_{k+2} = \Psi_{k+1}$
and/or $\Psi_{k-1} = \Psi_k$, difference the two values of $C_m$
obtained for $\psi_m = \Psi_{k+1} - \epsilon$ and $\psi_m = \Psi_k +
\epsilon$ in the limit that $\epsilon$ goes to zero.  In practice, the
computation is straightforward and it is not necessary to take limits.

The inverse computation, in which one determines a $\psi_m$ given a
coordinate $C_m$, is relatively straightforward, at least in the case
of a single separable coordinate.  One scans the coordinate vector
from the start looking for a pair of values that encompass $C_m$.
Having found a pair, one must then check the indexing vector.  If the
two index values are unequal, then the correct pair has been found.
Otherwise the search must be continued.  When a correct coordinate
pair has been found, the inverse of Eq.~(\ref{eq:tabcoord}) is
applied to determine $\Upsilon_m$.  The inverse of
Eq.~(\ref{eq:indxcoord}) then yields $\psi_m$.

It is understood that the general keywords \keyi{CUNIT}{m\Ci},
\keyi{CRDER}{m\Ci}, and \keyi{CSYER}{m\Ci} apply to the output
coordinate $C_m$ rather than the {\tt -TAB} coordinate keywords
\CRVAL{i\Ci} et al\@.  Similarly, if the table lookup determines
celestial coordinates, the general keywords \keyi{RADESYS}{\Ci} and
\keyi{EQUINOX}{\Ci} apply to the output celestial coordinates rather
than the input {\tt -TAB} coordinates.

\subsubsection{{\tt -TAB} implementation, parameters, and
requirements}

Standard FITS binary tables extensions ({\tt XTENSION = 'BINTABLE'})
will be used to hold the coordinate array in a single cell of a column
of a one-row table.  The length of this array is given in the FITS
table header by the repeat count in the \keyi{TFORM}{n} keyword, where
$n$ is the number of the column containing the coordinate array.
The dimensions of the coordinate array will be given in the FITS table
header by the keyword \keyi{TDIM}{n} set to {\tt '($M$, $K_1$, $K_2$,
$\ldots$ $K_M$)'}, where $n$ is the column containing the coordinate
array.  Note in particular that if $M = 1$ the coordinate array is
formally {\em two-dimensional} though the first axis is degenerate.
The repeat count in the \keyi{TFORM}{n} keyword value is the product
of $M$ and all the $K_m$.  The indexing vectors for each of the
$M$ axes, if present, will occupy separate columns, each containing a
one-dimensional array of length $K_m$. 

The {\tt BINTABLE} extension containing the coordinate array must be
in the same FITS file as the data that reference it.

The parameters required by {\tt -TAB} are the table extension name
({\tt EXTNAME}), the table version number ({\tt EXTVER}), the table
level number ({\tt EXTLEVEL}), the column name for the coordinate
array ({\tt TTYPE$n_1$}), the column name for the indexing vector
({\tt TTYPE$n_2$}), and the axis number $m$ associated with
intermediate world coordinate axis $i$ in the coordinate array.  The
keywords used for this purpose are \PS{i}{{\tt 0}\Ci},
\PV{i}{{\tt 1}\Ci}, \PV{i}{{\tt 2}\Ci}, \PS{i}{{\tt
1}\Ci}, \PS{i}{{\tt 2}\Ci}, and \PV{i}{{\tt 3}\Ci},
respectively.  These are summarized in Table~\ref{ta:TABkeys}. 
For images, \PS{i}{{\tt 0}\Ci}\ has no default; for tables a
missing or blank \PS{i}{{\tt 0}\Ci}\ is taken to be the current
table (see below).  \PV{i}{{\tt 1}\Ci}\ and \PV{i}{{\tt
2}\Ci}\ both have a default value of 1.  \PS{i}{{\tt 1}\Ci}\
never has a default; it must be present and be assigned a value
actually occurring in table \PS{i}{{\tt 0}\Ci}\@.  If 
\PS{i}{{\tt 2}\Ci}\ is missing or has a value of all blanks, the
indexing vector is taken to be a list of integers from 1 to $K_m$
and $\psi_m$ becomes a direct index of axis \PV{i}{{\tt 3}\Ci} in
the array specified by \PS{i}{{\tt 1}\Ci}\@.  If \PV{i}{{\tt
2}\Ci} is present and assigned a value, then that value must
actually occur in table \PS{i}{{\tt 0}\Ci}\@.  Note that the
values given to \PS{i}{{\tt 1}\Ci} and \PS{i}{{\tt 2}\Ci} are not case
sensitive since the FITS Standard (Hanisch et al.~\cite{kn:NOST})
states that ``String comparisons with the values of \keyi{TTYPE}{n}\
keywords should not be case sensitive.''

\begin{table}
\centering
  \caption{Parameter keywords used for the {\tt -TAB} algorithm.}
  \renewcommand{\arraystretch}{1.25}
  \begin{tabular}{llc}
     \hline\hline
     Keyword & Use & Default \\
     \hline
     \PS{i}{{\tt 0}\Ci} & table extension name ({\tt EXTNAME}) & --- \\
     \PV{i}{{\tt 1}\Ci} & table version number ({\tt EXTVER}) & 1 \\
     \PV{i}{{\tt 2}\Ci} & table level number ({\tt EXTLEVEL}) & 1 \\
     \PS{i}{{\tt 1}\Ci} & column name for the coordinate \\
                     & \hspace{1em} array ({\tt TTYPE$n_1$}) & --- \\
     \PS{i}{{\tt 2}\Ci} & column name for the indexing          & \\
                     & \hspace{1em} vector ({\tt TTYPE$n_2$}) &
                                                       no index \\
     \PV{i}{{\tt 3}\Ci} & axis number ($m$) in array \PS{i}{{\tt
                        1}\Ci}  & 1 \\ 
     \hline
   \end{tabular}
  \label{ta:TABkeys}
\end{table}

The use of {\tt -TAB} for $M$ related axes requires the header to
specify the array \PS{i}{{\tt 1}\Ci} to be the same for each of
the $M$ axes.  The dimensions of this array must be given in the
header as $(M, K_1, K_2, \ldots, K_M)$.  These dimensions determine
the maximum range of the array index (1 to $K_m$) for each axis.  The
indexing vectors for each of the $M$ axes, if present, will occupy
separate columns, each containing a one-dimensional array of length
$K_m$.  The $M$ values of \PV{i}{{\tt 3}\Ci} must account for all
$M$ axes. If any of these conditions are not met, the result is
undefined.

If a FITS file contains multiple {\tt XTENSION} HDUs (header-data
units) with the specified {\tt EXTNAME}, {\tt EXTLEVEL}, and 
{\tt EXTVER}, then the result of the WCS table lookup is undefined.
If the specified FITS {\tt BINTABLE} contains no column, or multiple
columns, with the specified \keyi{TTYPE}{n}, then the result of the
WCS table lookup is undefined.  The specified FITS {\tt BINTABLE} must
contain only one row.

No units conversions are to be performed. \CUNIT{i}\ must be the same
as {\tt TUNIT$n$} of the binary table, where $n$ is the column number
corresponding to \PS{i}{{\tt 1}\Ci}.

We recommend strongly that the value chosen for {\tt EXTNAME}
always begin with the four letters {\tt WCS-}\@.  If this is done,
generic programs will recognize the table as part of the WCS portion
of the data and will be less likely to separate the table from the
rest of the WCS\@.

The {\tt -TAB} implementation is more complicated than most other
WCS conventions because the coordinate system is not completely
defined by keywords in a single FITS header.  Software that supports
{\tt -TAB} must be able to gather all the necessary WCS parameters
that are in general distributed over two FITS HDUs and in the
body of the WCS extension table.  The producers of FITS data products
should consider the capabilities of the likely recipients of their
files when deciding whether or not to use the {\tt -TAB} convention,
and in general should use it only in cases where other simpler WCS
conventions are not adequate.

\subsubsection{{-TAB} usage in tables}
\label{s:TABtable}

Binary table extensions containing array data columns may need a
table-lookup function for coordinate values.  Seemingly the most
convenient form of {\tt -TAB} would be one in which the {\tt
-TAB} array(s) are also table cells in the same row as the data
array.  However, a separate coordinate table would be more
economical if it applied to data arrays in multiple rows. The
{\tt -TAB} keywords for such tables are {\it i\/}{\tt CTYP}{\it
n\Ci}, {\it i\/}{\tt S{\it n}\_0\Ci}, {\it i\/}{\tt
V{\it n}\_1\Ci}, {\it i\/}{\tt V{\it n}\_2\Ci}, {\it
i\/}{\tt S{\it n}\_1\Ci}, and {\it i\/}{\tt S{\it n}\_2\Ci},
where $n$ is the column number of the array of data and $i$ is
the intermediate world coordinate axis number.  If {\it i\/}{\tt
S{\it n}\_0\Ci} is missing or blank, the present binary table is
taken to be the coordinate lookup table.  In that case, the coordinate
array(s) are taken to be single-cell arrays in the same row
as the data array and keywords {\it i\/}{\tt V{\it
n}\_1\Ci} and {\it i\/}{\tt V{\it n}\_2\Ci} are ignored.

Strictly speaking, the {\tt -TAB} representation is not required for
binary or ASCII table extensions containing only one data value per
cell because each of the coordinate values associated with the datum
may be stored in separate columns.  However, {\tt -TAB} does provide a
convenient method of solving one of the problems pertaining to
such tables, namely identifying the data column.  If the table
contains a column of data and another one or more columns of
coordinate values pertaining to those data, then one can define the
coordinates of the data column as being of type {\tt -TAB}\@.  In this
case, the critical keywords are {\it i\/}{\tt CTYP}{\it n}\Ci\ to
declare the coordinate axis type and {\it i\/}{\tt  S{\it n}\_1\Ci} to
identify the corresponding coordinate column. Since the data value
column contains only one value, the coordinate column should only
contain one value.  Then, the usual coordinate keywords ({\it j\/}{\tt
CRPX}{\it n} / {\it j\/}{\tt CRP}{\it n\Ci},  {\it ij\/}{\tt PC}{\it
n\Ci}, {\it ij\/}{\tt CD}{\it n\Ci},  {\it i\/}{\tt CDLT}{\it n} /
{\it i\/}{\tt CDE}{\it n\Ci}) may be omitted since their defaults
yield the desired result that output index pixel equals the input data
pixel (both of which are 1 in the assumed case).  Note too that tables
of this type --- one value per table cell --- may be in either ASCII
or binary table form.


\subsection{{\tt -TAB} examples}

To illustrate the use of {\tt -TAB} with two non-separable axes,
let
us consider a specific example of a four-dimensional image array.
Assume that axes 1 and 3 are handled either linearly or by one of the
non-linear single-axis cases (including {\tt -TAB})\@.  However,
assume that axes 2 and 4 require table lookup in a mutually dependent
fashion.  Thus $i = 2$ for $m = 1$ and $i = 4$ for $m = 2$ and
the required keywords are\\
\begin{tabular}{l}
\verb!PS2_0 = 'WCS-TAB'     PS4_0 = 'WCS-TAB'!\\
\verb!PV2_1 = 1             PV4_1 = 1       !\\
\verb!PV2_2 = 1             PV4_2 = 1       !\\
\verb!PS2_1 = 'COORDS'      PS4_1 = 'COORDS'!\\
\verb!PS2_2 = 'INDEX2'      PS4_2 = 'INDEX4'!\\
\verb!PV2_3 = 1             PV4_3 = 2!
\end{tabular}\\
where the first four keyword pairs must match exactly and the last two
pairs must have different values.

A real example that could use this particular algorithm is
represented by a spectral image of a portion of the sky taken with a
single radio telescope.  The observer commands the telescope to point
at a regular grid of coordinates, but, due to wind loading and other
pointing errors, the telescope achieves the commanded positions only
approximately.  The actual celestial coordinates observed are,
however, accurately measured.  These data could then be represented
usefully as a two-dimensional array of spectra, but accurate celestial
coordinates for each spectrum could be found only by a 2-dimensional
table lookup.

Since there has been no generally agreed upon FITS format for
spectral data with explicit wavelengths assigned to each pixel, data
providers have resorted to defining their own formats.  Examples
from the Hubble Space Telescope and the Very Large Array are shown
below, recast into the {\tt -TAB} algorithm.


\subsubsection{{\tt -TAB} examples: HST data}

Two types of Hubble Space Telescope data serve as examples of the
{\tt -TAB} algorithm.  The two cases illustrate the evolution from
simple images to the more powerful constructs provided by FITS
extensions and binary tables.  The purpose of discussing these formats
is not to explain the HST formats, but to illustrate a couple of types
of data that can be represented by the {\tt -TAB} algorithm.
Therefore, some details of these formats are ignored.

The early HST spectrographs, FOS and GHRS, adopted a format based only
on simple FITS images.  The basic concept is that the spectral flux
values are given in one image and the vacuum wavelengths, in
Angstroms, are given as the pixel values in another image.  The two
are associated by filenames.  The filenames have the same basename,
but different filename extensions for an exposure.  To represent these
spectra with a FITS WCS based on the {\tt -TAB} method, while
continuing to use an image representation for the spectral fluxes,
one replaces the separate wavelength image with a table extension.

\begin{table}
\caption{WCS keywords in the spectral image for FOS and GHRS example.}
\renewcommand{\arraystretch}{1.05}
\setlength\tabcolsep{4.71pt}
\begin{tabular}{l}
\hline
\hline
\verb!CTYPE1  = 'WAVE-TAB'   / Coordinate type! \\
\verb!CUNIT1  = 'Angstrom'   / Coordinate units! \\
\verb!PS1_0   = 'WCS-TAB'    / Coord table extension name! \\
\verb!PS1_1   = 'WAVELENGTH' / Coord table column name! \\
\hline
\end{tabular}
\label{ta:HSTFOS}
\end{table}

\begin{table}
\caption{WCS keywords in the STIS table for two of the spectral
    columns.}
\renewcommand{\arraystretch}{1.05}
\setlength\tabcolsep{4.71pt}
\begin{tabular}{l}
\hline
\hline
\verb!1CTYP4  = 'WAVE-TAB'   / Coordinate type! \\
\verb!1CUNI4  = 'Angstrom'   / Coordinate units! \\
\verb!1S4_1   = 'WAVELENGTH' / Coords for gross spectrum! \\
\verb! ! \\
\verb!1CTYP5  = 'WAVE-TAB'   / Coordinate type! \\
\verb!1CUNI5  = 'Angstrom'   / Coordinate units! \\
\verb!1S5_1   = 'WAVELENGTH' / Coords background spectrum! \\
\hline
\end{tabular}
\label{ta:HSTSTIS}
\end{table}

Table~\ref{ta:HSTFOS} shows the minimum WCS keywords required in
the primary image header.  The \keyw{PS1\_0} keyword value defines the
coordinate table to be in a binary table extension with name {\tt
'WCS-TAB'}\@.  The coordinate table extension consists of one column,
named {\tt 'WAVELENGTH'}, and one row.  The array values and length
correspond to the original image format.

Other WCS keywords defined in Paper I would also be included as
needed.  The intermediate coordinate transformation must produce
indexing values equal to the pixel coordinate in the image.  Since the
defaults for these keywords are defined in Paper I to produce pixel
coordinates, it is not strictly necessary to include these keywords.

The second generation HST spectrograph, STIS, adopted a binary table
format very close to one of those defined for the {\tt -TAB}
algorithm.  In the STIS one-dimensional extracted echelle format each
exposure is stored in a separate binary table.  The extracted data
from an exposure consist of a number of echelle orders each of which
is stored in a separate row.  The wavelengths and various associated
``spectra'' are array elements in different columns.  An associated
spectrum is an array of a single type data quantity such as fluxes,
errors, or data quality flags.

To convert this format to the {\tt -TAB} representation only requires
adding the \keyi{1CTYP}{n} and \keyii{1S}{n}{\tt 1} keywords to
specify the coordinate type and coordinate column name for each
spectral column {\it n\/}.  There are six types of spectral quantities
for each order: gross, background, net, flux, error, and data quality
flag.  So there must be six pairs of keywords.  Since the same
wavelength column applies to each of the spectra, the keyword values
are repeated six times, but with different column numbers in the
keywords.  Table~\ref{ta:HSTSTIS} shows the WCS keywords for two
of the spectral columns.  The \keyii{1S}{n}{0} keywords are
omitted, signifying that the coordinate array column is in the same
table as the spectra.

As noted for the FOS and GHRS format, additional WCS keywords are
included as needed and the intermediate coordinate transformation
values may be absent since this defines an indexing by pixel
coordinate.  In the STIS data, the units of the wavelength array are
vacuum Angstroms so the \keyi{1CUNI}{n} keyword would be required.


\subsubsection{{\tt -TAB} example: radio interferometry}

\label{s:TABradio}

\begin{table}
\centering
\caption[]{Sample coordinate table.  Each displayed column is actually
a one-cell array of the (single-row) table.}
\renewcommand{\arraystretch}{1.25}
\begin{tabular}{cc}
\hline\hline
{\it i\/}{\tt S{\it n}\_2} & {\it i\/}{\tt S{\it n}\_1} \\
\hline
\noalign{\smallskip}
$ 1 $                     & $\nu_1 $                      \\
$ N_1 $                   & $\nu_1 + (N_1 - 1)\, \delta_1 $ \\
$ N_1 + 1 $               & $\nu_2 $                      \\
$ N_1 + N_2 $             & $\nu_2 + (N_2 - 1)\, \delta_2 $ \\
$ N_1 + N_2 + 1 $         & $\nu_3 $                      \\
$ N_1 + N_2 + N_3$        & $\nu_3 + (N_3 - 1)\, \delta_3 $ \\
$\ldots$                  & $\ldots$                      \\
$ \sum_1^{L-1}{N_\ell} + 1 $ & $\nu_L $                      \\
$ \sum_1^L{N_\ell} $         & $\nu_L + (N_L - 1)\, \delta_L $ \\
\noalign{\smallskip}
\hline
\end{tabular}
\label{ta:UVfreqs}
\end{table}

Radio interferometry data are now best represented in binary tables
with each row representing the visibility at one time on one antenna
pair with a vector of data representing the complex values for all
polarizations and frequencies observed.  These telescopes often allow
the user to observe $N_\ell$ regularly-sampled spectral channels
starting from $L$ arbitrarily chosen intermediate frequencies (IFs).
Furthermore, this spectral data pattern may be used with one receiver
and, a few minutes later, with other receivers at very different
frequencies.  Because of the great flexibility in the choice of
the IFs and receivers, the frequencies present in the data may only be
described by a table. Because that description is very repetitive, we
choose to put it in a table separate from the visibility table.

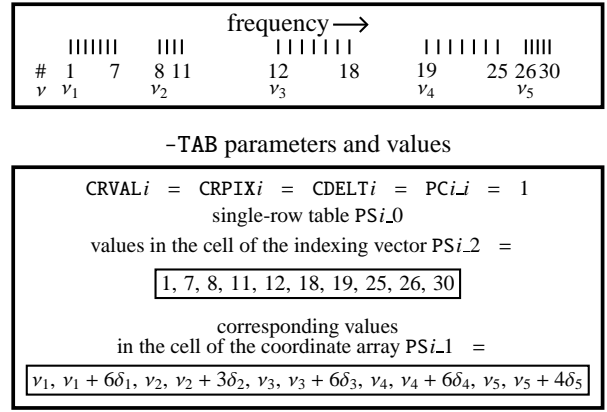
\begin{figure}
  \centerline{
    \resizebox{8cm}{!}{ \Large
      \begin{pspicture}(400pt,322pt)
        \psset{xunit=1pt,yunit=1pt,runit=1pt}
        \psset{linewidth=2.5}
        \rput(200,310){\LARGE multiple spectral channels}
        \rput(200,290){\LARGE at each of multiple base frequencies}
        \psframe(0,203)(400,275)
        \psframe(0,0)(400,165)
        \psset{linewidth=1.4}
        \rput(180,260){\LARGE frequency}
        \qline(217,260)(240,260)
        \psline(235,265)(240,260)(235,255)
        \qline( 40,250)(40,240)
        \qline( 45,250)(45,240)
        \qline( 50,250)(50,240)
        \qline( 55,250)(55,240)
        \qline( 60,250)(60,240)
        \qline( 65,250)(65,240)
        \qline( 70,250)(70,240)
        \qline(100,250)(100,240)
        \qline(105,250)(105,240)
        \qline(110,250)(110,240)
        \qline(115,250)(115,240)
        \qline(180,250)(180,240)
        \qline(188,250)(188,240)
        \qline(196,250)(196,240)
        \qline(204,250)(204,240)
        \qline(212,250)(212,240)
        \qline(220,250)(220,240)
        \qline(228,250)(228,240)
        \qline(280,250)(280,240)
        \qline(288,250)(288,240)
        \qline(296,250)(296,240)
        \qline(304,250)(304,240)
        \qline(312,250)(312,240)
        \qline(320,250)(320,240)
        \qline(328,250)(328,240)
        \qline(347,250)(347,240)
        \qline(351,250)(351,240)
        \qline(355,250)(355,240)
        \qline(359,250)(359,240)
        \qline(363,250)(363,240)
        \rput(20,229){\#}
        \rput(20,215){$\nu$}
        \rput(40,229){1}
        \rput(40,215){$\nu_1$}
        \rput(70,229){7}
        \rput(100,229){8}
        \rput(100,215){$\nu_2$}
        \rput(115,229){11}
        \rput(180,229){12}
        \rput(180,215){$\nu_3$}
        \rput(228,229){18}
        \rput(280,229){19}
        \rput(280,215){$\nu_4$}
        \rput(328,229){25}
        \rput(347,229){26}
        \rput(347,215){$\nu_5$}
        \rput(363,229){30}
        \rput(200,178){\LARGE{\tt -TAB} parameters and values}
        \rput(200,150){\keyi{CRVAL}{i}\ \eq\ \keyi{CRPIX}{i}\ \eq\
            \keyi{CDELT}{i}\ \eq\ \keyii{PC}{i}{i}\ \eq\ 1}
        \rput(200,130){single-row table \keyii{PS}{i}{0}}
        \rput(200,110){values in the cell of the indexing vector
             \keyii{PS}{i}{2}\ \eq}
        \rput(200,85){\psframebox{$1 ,\, 7 ,\, 8 ,\, 11 ,\, 12 ,\,
              18 ,\, 19 ,\, 25 ,\, 26 ,\, 30$}}
        \rput(200,55){corresponding values}
        \rput(200,42){in the cell of the coordinate array
             \keyii{PS}{i}{1}\ \eq}
        \rput(200,20){\psframebox{$\nu_1,\, \nu_1+6\delta_1,\,
             \nu_2,\, \nu_2+3\delta_2,\, \nu_3,\, \nu_3+6\delta_3,\,
             \nu_4,\, \nu_4+6\delta_4,\, \nu_5,\, \nu_5+4\delta_5$}}
      \end{pspicture}
    }
  }
\caption[]{Example taken from radio interferometry using {\tt
-TAB} with an indexing vector.  The FITS keywords shown are suitable
for the random groups format.  The observation is made at a number of
frequencies, with a number $N_\ell$ of regularly spaced $\delta_\ell$
spectral channels beginning from each of a number of arbitrary
base frequencies $\nu_\ell$.  The use of an indexing vector reduces
the number of values in the table from one array of 30 to two arrays
of 10 each.  In a real case the number of spectral channels would be
significantly larger, making the space savings significant.  In this
example, pixel $p_j = 6$ produces $\psi_m = 6$.  This lies at
$\Upsilon_m = 1\frac{5}{6}$.  The resulting coordinate is then
$\frac{1}{6}\nu_1 + \frac{5}{6}\left( \nu_1 + 6\delta_1 \right)$
which, as one would expect, equals $\nu_1 + 5\delta_1$.  Note
that this example involves only a single independent {\tt -TAB} axis,
so that \PV{i}{{\tt 3}} $= m \equiv 1$.}
\label{fig:TABradio}
\end{figure}

We can construct a single column table of all
$\sum_{\ell=1}^L{N_\ell}$ frequencies observed.  However, by using a
second column for an indexing vector, we can take advantage of the
regular sampling of the $N_\ell$ spectral channels.  If we set {\it
i\/}{\tt CRVL}{\it n}, {\it i\/}{\tt CDLT}{\it n}, and {\it i\/}{\tt
CRPX}{\it n} (for binary table format) or \keyi{CRVAL}{i},
\keyi{CDELT}{i}, and \keyi{CRPIX}{i} (for random groups and image
formats) to one, then Table~\ref{ta:UVfreqs} may represent the actual
frequencies, where $\nu_\ell$ is the frequency of channel 1 in IF
$\ell$ and $\delta_\ell$ is the increment in frequency between the
regularly sampled channels of IF $\ell$.  This example is shown
graphically in Fig.~\ref{fig:TABradio}.

\subsubsection{{\tt -TAB} examples: multiple exposures and sampling}
\label{s:Steve}

In general, the FITS WCS papers do not consider the process by which
the physical effects represented by the values in the FITS array have
been measured.  The measurable quantity may have been sampled over
regions small enough to be considered as points in a widely separated
grid of world coordinates.  Alternatively, the measurable quantity may
have been integrated across overlapping adjacent regions.  The FITS
WCS papers do not indicate whether it is valid to presume that the
values in adjacent array elements represent physically adjacent
quantities, nor do they indicate whether it is valid to interpolate
the WCS across the extent of an array element.

Some world coordinate axes are defined to have integral values with
conventional meanings (e.g.~{\tt COMPLEX}, {\tt STOKES}).
Interpolating a world coordinate value across the extent of a FITS
array element along the direction of an axis which is conventionally
defined to be integral would clearly be inappropriate.  Nevertheless,
for other types of world coordinate axes in traditional FITS arrays,
the adjacent elements in the {\tt NAXIS}-dimensional array can often
be presumed to represent adjacent locations in a measurement
continuum.  This adjacency is relevant, for example, to image display
programs when the image is magnified so that one FITS array element
extends across multiple display pixels.  In such cases a FITS display
program may attempt to interpolate WCS values across the extent of an
array element.  With \keyi{CTYPE}{k} values other than {\tt -TAB} this
is commonly done by linear interpolation between the presumably
adjacent array elements.  However for coordinate axes where the
\keyi{CTYPE}{k} is described by the {\it xxxx\/}{\tt -TAB} non-linear
algorithm there can be no presumption of adjacency.

\begin{table*}
\caption{Top: keywords in main FITS header for {\tt -TAB} example,
bottom: table keywords and sample arrays for that example.}
\centering
\begin{tabular}{l}
\hline
\hline
\noalign{\smallskip}
\verb+123456789 123456789 123456789 123456789 123456789 123456789 123456789 123456789 + \\
\noalign{\smallskip}
\hline
 \noalign{\smallskip}
\verb!NAXIS   =                    3 / 2-d celestial coord, 1-d multiple exposure! \\
\verb!NAXIS3  =                    4 / radio, infrared, optical, x-ray! \\
\verb!WCSAXES =                    4 / multiple exposures in bandpass and time! \\
\verb!WCSNAME = 'Multi-wavelength, multi-epoch'! \\
\verb!CTYPE3  = 'WAVE-TAB'           / spectral axis by table look-up! \\
\verb!CTYPE4  = 'TIME-TAB'           / temporal axis by table look-up! \\
\verb!CNAME3  = 'Bandpass'           ! \\
\verb!CNAME4  = 'Observation Date'   ! \\
\verb!CUNIT3  = 'm'                  / wavelengths in meters! \\
\verb!CUNIT4  = 'a'                  / observation dates in years! \\
\verb!CRPIX3  =                  0.5 / spectral bin initial edge of initial pixel! \\
\verb!CRPIX4  =                    1 / temporal bins are at begin/end time! \\
\verb!CDELT3  =                    1 / spectral index array is one unit per bin! \\
\verb!CDELT4  =                    1 / temporal index array is one unit per bin! \\
\verb!CRVAL3  =                  0.5 / spectral reference is center of radio exposure! \\
\verb!CRVAL4  =                    0 / temporal reference is start of radio exposure! \\
\verb!PC3_3   =                    1 / 3rd pixel axis increments 3rd image coord axis! \\
\verb!PC3_4   =                    0 ! \\
\verb!PC4_3   =                    1 / 3rd pixel axis increments 4th image coord axis ! \\
\verb!PC4_4   =                    1 / 4th pixel axis is degenerate (1 point)! \\
\verb!PS3_0   = 'WCS-table'! \\
\verb!PS3_1   = 'WaveCoord'! \\
\verb!PS3_2   = 'WaveIndex'! \\
\verb!PS4_0   = 'WCS-table'! \\
\verb!PS4_1   = 'TimeCoord'! \\
\verb!PS4_2   = 'TimeIndex'! \\
\verb! ! \\
\hline
\verb! ! \\
\verb!EXTNAME = 'WCS-table'! \\
\verb!EXTVER  =                    1 ! \\
\verb!EXTLEVEL=                    1 ! \\
\verb! ! \\
\verb!                        TUNIT2 = 'm'                                    TUNIT4 = 'a'! \\
\verb!TFORM1 = '8E'           TFORM2 = '8D'           TFORM3 = '8E'           TFORM4 = '8D'! \\
\verb!TTYPE1 = 'WaveIndex'    TTYPE2 = 'WaveCoord'    TTYPE3 = 'TimeIndex'    TTYPE4 = 'TimeCoord'! \\
\verb!--------------------    --------------------    --------------------    --------------------! \\
\verb! (0.5,                   (0.21106114,            (0.0,                   (1997.84512,! \\
\verb!   1.5,                    0.21076437,             1.0,                    1997.84631,! \\
\verb!    1.5,                    2.0E-6,                 1.0,                    1993.28451,! \\
\verb!     2.5,                    2.2E-6,                 2.0,                    1993.28456,! \\
\verb!      2.5,                    500.0E-9,               2.0,                    2001.59234,! \\
\verb!       3.5,                    650.0E-9,               3.0,                    2001.59239,! \\
\verb!        3.5,                    1.24E-9,                3.0,                    2002.18265,! \\
\verb!         4.5)                    2.48E-9)                4.0)                    2002.18301)! \\
\noalign{\smallskip}
\hline
\end{tabular}
\label{ta:widths}
\end{table*}

Notwithstanding these caveats, the use of {\tt -TAB} does permit
specification of changes in the coordinate value across the extent of
a single array element.  Consider a sequence of two-dimensional images
of the sky which have been re-gridded such that the pixel array for
each image shares the same 2-D WCS for celestial coordinates.  A
single 3-D FITS array can hold such a sequence of 2-D images.  If the
same instrument produced each of the images on different dates then
the third dimension might be purely temporal.  Alternatively, if
different instruments (radio, IR, optical, X-ray) produced each of the
images then the third dimension might be spectral.  In practice,
however, images in such a spectral 3-D image will typically also have
different observation dates, so the image will actually span four
dimensions of world coordinates.

Akin to the example of a long slit spectrum in Paper II, the WCS of
this four-dimensional case can be represented by the 3-D array.  This
fourth dimension could be represented with {\tt NAXIS = 4} and {\tt
NAXIS4 = 1}, or with {\tt NAXIS = 3} and {\tt WCSAXES = 4}.  Whether
the sequence of images is temporal or spectral, it is certain that the
exposure duration or the spectral bandpass have a non-zero extent and
the {\tt -TAB} representation does provide the ability to communicate
the start and end times of an observation, or the minimum and maximum
wavelengths of an observation. For this example the WCS keywords are
shown in Table~\ref{ta:widths}, where the keywords relating to axes 1
and 2 are omitted because they are simply celestial coordinates as
described by Paper II\@.  In this example, note the clever use of
the {\tt PC} matrix to cause the fourth coordinate axis to depend on
the third pixel axis, but not the degenerate fourth pixel axis.  FITS
writers can use details such as these to communicate how a display
program might provide meaningful world coordinates at locations across
an array element.\footnote{The precise interpretation of
\keyi{CTYPE}{k} keywords that describe {\tt TIME} is deferred to a
possible future paper on temporal coordinates in FITS\@.}

\begin{table*}
\centering
\caption[]{Recognized values for {\tt SPECSYS}\Ci, {\tt
    SSYSOBS}\Ci, and {\tt SSYSSRC}\Ci.}
   \renewcommand{\arraystretch}{1.25}
\protect\begin{tabular}{lllrrl}
\hline\hline
Value & Definition & Magnitude & $\ell$ & $b$ & Reference \\
\hline
{\tt TOPOCENT} &Topocentric  &  0.0\,km\,\,s$^{-1}$
               & -- & -- & \\
{\tt GEOCENTR} &Geocentric   &  0.5
               & -- & -- & \\
{\tt BARYCENT} &Barycentric & 30
               & -- & -- & Stumpff~(\cite{kn:S}) \\
{\tt HELIOCEN} &Heliocentric & 30
               & -- & -- & Stumpff~(\cite{kn:S}) \\
{\tt LSRK}     &Local standard of rest (kinematic) & 20
               & $\dag$ 56 & 23
               & Delhaye~(\cite{kn:D}), Gordon~(\cite{kn:G}) \\
{\tt LSRD}     &Local standard of rest (dynamic) & 16.6
               & 53 & 25 & Delhaye~(\cite{kn:D}) \\
{\tt GALACTOC} &Galactocentric & 220
               & 90 & 0 & Kerr \& Lynden-Bell~(\cite{kn:KLB})\\
{\tt LOCALGRP} &Local Group & 300
               & 90 & 0 & de Vaucouleurs~(\cite{kn:dV})\\
{\tt CMBDIPOL} & Cosmic microwave background dipole & 368
               & 263.85 & 48.25
               & Bennett et al.~(\cite{kn:BHHJ}) \\
{\tt SOURCE}   &Source rest frame & any
               & -- & -- &  \\
\hline
\multicolumn{6}{l}{$\dag$ {\tt LSRK} is normally quoted as right
    ascension 18 hours, declination 30 degrees (1900).}
\end{tabular}
\label{ta:specsys}
\end{table*}

In order to provide coordinate values across the entire array element,
the coordinate values on the boundaries between the array elements are
multiply defined.  Such a result should be expected for array elements
representing non-adjacent physical values.  It is evident in this case
that the order of the index arrays should not be permuted during the
coordinate lookup.  The lower portion of Table~\ref{ta:widths} shows a
subset of the content of the table(s) referred to by \PS{{\tt 3}}{{\tt
0}} and \PS{{\tt 4}}{{\tt 0}}.  The four columns in the figure are
meant to illustrate the association of table header keywords and table
values.  In the FITS file, the keywords would appear in the table
header along with many other keywords while the four arrays of
values would appear in four cells of the one-row binary table.

The indexing vectors in the table of Table~\ref{ta:widths}
illustrate the reason to allow two values within the indexing
vector to be identical. The rules for the linear interpolation within
such vectors are clear.  One must select the two index locations
having values immediately surrounding $\psi_m$.  Thus in the present
example, if the time coordinate has $\psi_m = 1.1$, the third and
fourth index and coordinate values are interpolated.  The output time
coordinate is 
\begin{displaymath}
    1993.28451 \, + \,\frac{1.1 - 1.0}{2.0 - 1.0} \,\,
        (1993.28456 - 1993.28451)  \cdot 
\end{displaymath}
The result of the table look up is undefined if the $\psi_m$ is
equal to the repeated value in the indexing vector.



\section{Coordinate reference frames}
\label{s:reference}

Frequencies, wavelengths, and apparent radial velocities are
always referred to some selected standard of rest (reference frame),
and while they are measured, of necessity, in the observer's rest
frame, they may also be corrected to some other
frame.\footnote{Proper relativistic formul\ae\ should be used;
see Gibbs \&\ Tao (\cite{kn:GT}) or Rindler (\cite{kn:R}).  Better
yet, observers should consider using standard software described by
Lindegren and Dravins (\cite{kn:LD}) to perform the conversion from
topocentric coordinate to the standardized Barycentric Celestial
Reference System.}  The velocity correction is computed from the
vector dot product of the direction vector and the relative velocity
vector of the two reference frames.  In addition, the corrections from
topocentric, the frame in which the observations are usually made, to
geocentric and then to barycentric or heliocentric are dependent on
the dot product with time-variable velocity vectors.  As a
consequence, the ``velocity'' with respect to the reference frame 
depends on direction.  Differential effects over a field of view may
be important in some high precision applications; for example, they
may amount to 5\,km\,\,s$^{-1}$ for the Local Group correction over
a field of one degree.  In another example, observations of Galactic
CO over two-degree fields separated by two degrees failed to align by
two spectral channels (Mangum et al.~\cite{kn:MEG}).

If a frequency, wavelength or apparent radial velocity associated
with an image plane has been corrected to some other frame by
transforming the \CRVAL{k\Ci}\ et al.~values, then differential errors
may arise at points away from the reference point.  For example, it is
normal in radio astronomy to observe a field while holding constant the
velocity of the reference point with respect to the local standard of
rest. See Sect.~\ref{s:real comp} for a more detailed discussion of
the complications inherent in such data.

Nonetheless, each image plane shares a constant topocentric frequency
(or apparent radial velocity).  The velocity ({\it and\/}
frequency and wavelength) with respect to the local standard of rest
is then a function of celestial coordinate within the image.  In order
to denote this, we introduce two character-valued keywords.  The
first,\begin{center}
\begin{tabular}{l}
\noalign{\vspace{-5pt}}
\keyw{SPECSYS\Ci} \hspace{2em} (character-valued)\\
\noalign{\vspace{-5pt}}
\end{tabular}
\end{center}
\noindent describes the reference frame in use for the spectral-axis
coordinate(s).  The second,\begin{center}
\begin{tabular}{l}
\noalign{\vspace{-5pt}}
\keyw{SSYSOBS\Ci} \hspace{2em} (character-valued)\\
\noalign{\vspace{-5pt}}
\end{tabular}
\end{center}
\noindent describes the spectral reference frame that is constant over
the range of the non-spectral world coordinates and has default value
{\tt 'TOPOCENT'}.  The recognized values are given in
Table~\ref{ta:specsys}.  In this table, the magnitude column gives the
approximate magnitude of the velocity vector that defines the
particular reference frame and the $\ell$ and $b$ columns give the
standard Galactic longitude and latitude of the reference
frame, Blaauw et al~(\cite{kn:Blaauw}).
This table includes the traditional heliocentric system which the
authors believe should be deprecated. It has frequently been misused
as an alias for barycentric and would, if high accuracy is desired,
require the observing date in order to allow conversions between it
and the other systems.  The table also includes the WMAP microwave
background dipole as a velocity system.  Since, as the data are
refined over time, this system will change, its use at present should
be restricted to situations for which it is particularly appropriate
and the parameters used should be well commented.   The
radio-astronomy example described above has {\tt SPECSYS = 'LSRK'} and
{\tt SSYSOBS = 'TOPOCENT'} indicating that the spectral axis is
expressed as a kinematic Local Standard of Rest velocity, but that
each right ascension by declination image plane is at a constant
topocentric spectral coordinate.

Many spectrometers suffer from a significant variation in their
spectral coordinate as a function of celestial coordinate.  Such
variations are the primary subject of Paper IV, while the concepts of
this work refer to images produced by, or corrected to, an ideal
spectrometer.  In general we recommend that the frame corrections
will get folded into the instrumental ones and that the image produced
after the Paper IV corrections will have identical {\tt SSYSOBS} and
{\tt SPECSYS}\@.  This is not required, but would certainly
simplify matters.

It is not appropriate for this work to define the parameters of
the recognized rest frames.  The parameters needed to compute
geocentric frequencies/velocities from topocentric are the sidereal
time (or Earth rotation angle) and the observatory location.  The
observing date is needed to convert from geocentric to barycentric
coordinates.  The new keywords
\begin{center}
\begin{tabular}{l}
\noalign{\vspace{-5pt}}
\keyw{MJD-AVG} \hspace{2em} (floating-valued)\\
\noalign{\vspace{-5pt}}
\end{tabular}
\end{center}
and
\begin{center}
\begin{tabular}{l}
\noalign{\vspace{-5pt}}
\keyw{DATE-AVG} \hspace{2em} (character-valued)\\
\noalign{\vspace{-5pt}}
\end{tabular}
\end{center}
\noindent are reserved to give a representative time for the whole
observation, suitable for calculating proper motion, apparent place,
local apparent sidereal time, velocity with respect to the center of
the Earth and barycenter, etc.  The {\tt DATE-AVG} keyword shall
be expressed in the manner described by Hanisch et
al.~(\cite{kn:NOST}) for the {\tt DATE-OBS} keyword.  If both keywords
are present, they must agree to adequate accuracy or the result is
undefined.  If both keywords are absent, {\tt DATE-OBS} may be used to
determine the representative time, although {\tt DATE-OBS} refers to
the beginning of an observation and hence is not entirely suitable.
For high-precision applications, a time system to apply to {\tt
MJD-AVG} must be defined in the manner proposed by Bunclark \&\ Rots
(\cite{kn:BR}).  Note also that, for long integrations, no single
time is adequate for computing proper motion, sidereal time, and the
like.

Astronomers have traditionally quoted telescope positions using
terrestrial longitude, latitude, and height, but, for computation of
topocentric velocity, these values are immediately converted into a
geocentric Cartesian triple.  A precise conversion from longitude,
latitude, and height, however, requires knowledge of the geodetic
datum in which they are expressed.  There are about 1000 geodetic
datums in use.  These are described by the United States Department
of Defense (\cite{kn:DoD}) in a printed document also available over
the Internet.  Some of the datums deviate from geocentricity by over a
kilometer.

\begin{table*}
\caption[]{New spectral keywords including forms for use in tables.
All table keywords except the coordinate name have the same form
for the {\tt BINTABLE} vector image format and the pixel list
format.}
\centering
   \renewcommand{\arraystretch}{1.25}
\begin{tabular}{llllcl}
\hline\hline
Primary & BINTABLE & Pixel & Type & Sect.& Keyword\rule{0mm}{4mm}   \\
Array   &          & list  &      &      & Description \rule[-2mm]{0mm}{3mm}    \\
\hline
\keyi{CNAME}{i\Ci}&{\it i\/}{\tt CNA}{\it n\Ci} & {\tt TCNA}{\it n\Ci}
                  &character&\ref{s:CNAME}
                  &Coordinate axis name \\
{\tt RESTFRQ}\Ci  &\multicolumn{2}{c}{{\tt RFRQ{\it n\Ci}}}
                  &floating &\ref{s:COPAR}
                  &Line rest frequency (Hz)\\
{\tt RESTFREQ}    &\multicolumn{2}{c}{---}
                  &floating &\ref{s:COPAR}
                  &Line rest frequency alternate for primary array only\\
{\tt RESTWAV}\Ci  &\multicolumn{2}{c}{{\tt RWAV{\it n\Ci}}}
                  &floating &\ref{s:COPAR}
                  &Line rest wavelength in vacuum (m)\\
{\tt SPECSYS}\Ci  &\multicolumn{2}{c}{{\tt SPEC{\it n\Ci}}}
                  &character&\ref{s:reference}
                  &Spectral reference frame (from Table~\ref{ta:specsys})\\
{\tt SSYSOBS}\Ci  &\multicolumn{2}{c}{{\tt SOBS{\it n\Ci}}}
                  &character&\ref{s:reference}
                  &Spectral reference frame fixed during observation
                  (from Table~\ref{ta:specsys})\\
{\tt MJD-AVG}     &\multicolumn{2}{c}{{\tt MJDA{\it n}}}
                  &floating &\ref{s:reference}
                  &Average date of observation (Julian date$-2400000.5$) \\
{\tt DATE-AVG}   &\multicolumn{2}{c}{{\tt DAVG{\it n}}}
                  &character &\ref{s:reference}
                  &Average date/time of observation \\
{\tt OBSGEO-X}    &\multicolumn{2}{c}{{\tt OBSGX{\it n}}}
                  &floating &\ref{s:reference}
                  &Observation X (m) \\
{\tt OBSGEO-Y}    &\multicolumn{2}{c}{{\tt OBSGY{\it n}}}
                  &floating &\ref{s:reference}
                  &Observation Y (m) \\
{\tt OBSGEO-Z}    &\multicolumn{2}{c}{{\tt OBSGZ{\it n}}}
                  &floating &\ref{s:reference}
                  &Observation Z (m)\\
{\tt VELOSYS}\Ci  &\multicolumn{2}{c}{{\tt VSYS{\it n\Ci}}}
                  &floating &\ref{s:reference}
                  &Radial velocity wrt standard of rest (m\,s$^{-1}$)\\
{\tt ZSOURCE}\Ci  &\multicolumn{2}{c}{{\tt ZSOU{\it n\Ci}}}
                  &floating &\ref{s:reference}
                  &Redshift of source for {{\tt SOURCE}} cases (unitless)\\
{\tt SSYSSRC}\Ci  &\multicolumn{2}{c}{{\tt SSRC{\it n\Ci}}}
                  &character&\ref{s:reference}
                  &Spectral reference frame for {\tt SOURCE} cases
                  (from Table~\ref{ta:specsys})\\
{\tt VELANGL}\Ci  &\multicolumn{2}{c}{{\tt VANG{\it n\Ci}}}
                  &floating &\ref{appen:relativity}
                  &Angle of true velocity from tangent to line of
                   sight\rule[-2mm]{0mm}{3mm}\\
\hline
\end{tabular}
\label{ta:Vkeyword}
\end{table*}

Rather than burden FITS with these details of geodetic tradition,
three new keywords\begin{center}
\begin{tabular}{l}
\noalign{\vspace{-5pt}}
\keyw{OBSGEO-X} \hspace{2em} (floating-valued)\\
\keyw{OBSGEO-Y} \hspace{2em} (floating-valued)\\
\keyw{OBSGEO-Z} \hspace{2em} (floating-valued)\\
\noalign{\vspace{-5pt}}
\end{tabular}
\end{center}
\noindent are reserved to define a representative location for the
observation in a standard terrestrial reference frame.  The location
values should be right-handed, geocentric, Cartesian coordinates valid
at the epoch of {\tt MJD-AVG} or {\tt DATE-AVG}\@.  Position
errors of several kilometers should have negligible impact on the
computation of the diurnal velocity correction, so for this purpose
geodetic accuracy is not a requirement.  However, it should be
possible to provide coordinates with an accuracy of 1m or better based
on a recent satellite-based geodetic reference frame and we recommend
that FITS writers do so.  Although it is beyond the scope of this
paper to define particular terrestrial reference frames or tectonic
models, only frames based on the ITRS (McCarthy \&\ Petit
\cite{kn:ITRS}) are suitable for high-precision applications.  Most
post-1980 reference frames, including GPS, agree with recent versions
of the ITRF to about 0.1\,m.  Web sites\\
\hphantom{Mm}{\tt http://earth-info.nga.mil}/GandG/\\
\hphantom{Mm}{\tt http://www.ngs.noaa.gov/TOOLS/XYZ/xyz.html}\\
provide general geodetic information and a solution based
on the outputs of almost any hand-held GPS unit,
respectively.  These references, together with information about the
origin of the traditional telescope position, should allow the
determination of the keyword values defined in this paper.

It may be helpful for the FITS writer to provide information on the
relative radial velocity between the observer and the selected
standard of rest in the direction of the celestial reference
coordinate.  The keyword
\begin{center}
\begin{tabular}{l}
\noalign{\vspace{-5pt}}
\keyw{VELOSYS\Ci} \hspace{2em} (floating-valued)\\
\noalign{\vspace{-5pt}}
\end{tabular}
\end{center}
\noindent is hereby reserved for this purpose; its units shall be
m\,s$^{-1}$. \CUNIT{k\Ci}\ is not used since the WCS version \Ci\
might not be expressed in velocity units.

The frame of rest defined with respect to the source ({\tt SOURCE}) is
useful for comparing internal apparent radial velocities in
objects at different redshifts.  This allows the FITS writer to apply
all the cosmological and other model-dependent corrections,
leaving a coordinate local to the object.  For this frame of
rest, it is necessary to define the velocity with respect to some
other frame of rest.  The keywords 
\begin{center}
   \begin{tabular}{l}
      \noalign{\vspace{-5pt}}
      \keyw{ZSOURCE\Ci} \hspace{2em} (floating-valued)\\
      \noalign{\vspace{-5pt}}
   \end{tabular}
\end{center}
and
\begin{center}
   \begin{tabular}{l}
      \noalign{\vspace{-5pt}}
      \keyw{SSYSSRC\Ci} \hspace{2em} (character-valued)\\
      \noalign{\vspace{-5pt}}
   \end{tabular}
\end{center}
\noindent are hereby reserved; they should be used to document
the adopted value for this systemic velocity of the source.
{\tt ZSOURCE}\Ci\ is a unitless redshift; \keyw{SSYSSRC\Ci}
specifies the reference frame for this parameter and may have any
value in Table~\ref{ta:specsys} except {\tt SOURCE}\@.  The new
floating-valued and character-valued keywords hereby reserved are
listed\footnote{The frame and velocity keywords in
Table~\ref{ta:Vkeyword} apply to terrestrial observers only.
Observations from spacecraft require additional description of their
motions -- either a position and velocity or an ephemeris.  The
required keywords and/or tables for this are beyond the scope of this
paper.} in Table~\ref{ta:Vkeyword} along with the other keywords
defined in this paper.


\section{Alternate FITS image representations: pixel list and vector
column elements\protect\footnotemark}\footnotetext{Suggested by
contributions from William Pence, Arnold Rots, and Lorella Angelini of
the NASA Goddard Space Flight Center, Greenbelt, MD 20771.}

\label{s:bint}

The use of coordinate keywords to describe a multi--dimensional vector
in a single element of a FITS binary table (Cotton et
al.~\cite{kn:CTP}), and a tabulated list of pixel coordinates in a
FITS ASCII (Harten et al.~\cite{kn:HGGW}) or binary table was
discussed in Papers I and II\@.  In this section, we extend those
discussions to the keywords specific to spectral coordinates.  This
convention is considered an integral part of the full world
coordinates convention.


\subsection{Keyword naming convention}

Table~\ref{ta:Vkeyword} lists the corresponding set of coordinate
system keywords for use with each type of FITS image representation.
The keywords defined in this paper and their allowed values are
applicable to all three image storage formats.  The following notes
apply to the naming conventions used in Table~\ref{ta:Vkeyword}:

\begin{itemize}

\item
\Ci\ is a 1-character coordinate version code and may be blank
(primary) or any single uppercase character from {\tt A} through {\tt
Z}\@.

\item
$n$ is an integer table column number without any leading zeros
(1--999).

\item
$i$ is a one-digit integer axis number (1--9).
\end{itemize}

When using the {\tt BINTABLE} vector image format, if the table only
contains a single image column or if there are multiple image columns
but they all have the same value for any of the keywords in
Table~\ref{ta:Vkeyword} except \keyi{CNAME}{i\Ci}, then the
simpler form of the keyword name, as used for primary arrays, may be
used.  For example, if all the images in the table have the same
location, then one may use one set of {\tt OBSGEO-X}, {\tt OBSGEO-Y},
and  {\tt OBSGEO-Z} keywords rather than multiple \keyi{OBSGX}{n},
\keyi{OBSGY}{n}, and \keyi{OBSGZ}{n} keywords.  If both forms of
these keywords appear, then column-specific values are applied to
the specified columns and the non-specfic values are applied to all
other columns.  (Note, however, that the WCS keywords defined
for tables in Papers I and II must always be specified using the more
complex keyword name with the column number suffix and the axis number
prefix.)


\section{Spectral coordinate variation with other coordinates}
\label{s:spatial}

There are instruments having spectral coordinates that are a
function of other coordinates in the resulting image, e.g.~celestial
position. For example, optical astronomers use imaging Michelson
interferometers and scanning Fabry-Perot instruments to produce
three-dimensional images with two celestial and one spectroscopic
coordinate.  Because the path length through such instruments
increases off the optical axis of the telescope, the observed
wavelength is a function of celestial coordinate in each plane of the
image cube.  In general, this is a subject reserved for Paper IV since
such instruments almost always have lesser distortions that need
to be parameterized and corrected along with the path length
correction.

One particularly difficult case should be mentioned here.  Objective
prism instruments produce data in which the image of each object in
the field has been spread out into a spectrum.  Thus the output image
is the convolution of the sky with both a spatial
point-spread-function and a spectral dispersive point-spread function,
both in the two dimensions of the output image.  The methods used to
specify coordinates described in the present paper and in Papers
I and II are unable to handle such complex data. However, rather
than ignore the world coordinates entirely, writers of such images
should consider providing the celestial coordinates that would be
correct for the image if it had been taken at a single representative
wavelength.  That wavelength would be specified in a third WCS axis.
Alternate axis descriptions could then be used to specify the same
information for other wavelengths.  Multi-order echelle
spectrographs also produce images in which multiple spectral orders
overlap each other as well as the celestial axes and, hence, their
coordinates cannot be described by the methods of Papers II and
III\@.

Analysis of both objective prism and echelle spectrograph
images can produce tables of spectra that are easily described by
the present methods.


\section{Example}
\subsection{Basic spectral-line header}
\label{s:S-L_header}

\begin{table}
\caption[]{Example image header.}
\renewcommand{\arraystretch}{1.05}
\begin{tabular}{l}
\hline
\hline
\noalign{\smallskip}
\verb+123456789 123456789 123456789 123456789 123456789+ \\
\noalign{\smallskip}
\hline
\noalign{\smallskip}
\verb!SIMPLE  =                    T  ! \\
\verb!BITPIX  =                  -32  ! \\
\verb!NAXIS   =                    3  ! \\
\verb!NAXIS1  =                 1024  ! \\
\verb!NAXIS2  =                 1024  ! \\
\verb!NAXIS3  =                   63  ! \\
\verb!EXTEND  =                    T / Tables may follow! \\
\verb!OBJECT  = '3C353'               ! \\
\verb!TELESCOP= 'VLA'                 ! \\
\verb!DATE-OBS= '1998-09-29'         / Start obs! \\
\verb!BUNIT   = 'Jy/beam'             ! \\
\verb!EQUINOX =      2.000000000E+03  ! \\
\verb!DATAMAX =      5.080558062E-01  ! \\
\verb!DATAMIN =     -1.179690361E-01  ! \\
\verb!WCSNAME = 'TopoFreq'            ! \\
\verb!CTYPE1  = 'RA---SIN'            ! \\
\verb!CRVAL1  =    2.60108333333E+02  ! \\
\verb!CDELT1  =     -2.777777845E-04  ! \\
\verb!CRPIX1  =                512.0  ! \\
\verb!CUNIT1  = 'deg'! \\
\verb!CTYPE2  = 'DEC--SIN'            ! \\
\verb!CRVAL2  =   -9.75000000000E-01  ! \\
\verb!CDELT2  =      2.777777845E-04  ! \\
\verb!CRPIX2  =                513.0  ! \\
\verb!CUNIT2  = 'deg'! \\
\verb!CTYPE3  = 'FREQ'                ! \\
\verb!CRVAL3  =    1.37835117405E+09  ! \\
\verb!CDELT3  =      9.765625000E+04  ! \\
\verb!CRPIX3  =                 32.0  ! \\
\verb!CUNIT3  = 'Hz'! \\
\verb!OBSGEO-X=     -1.601185365E+06 / [m] ! \\
\verb!OBSGEO-Y=     -5.041977547E+06 / [m] ! \\
\verb!OBSGEO-Z=      3.554875870E+06 / [m] ! \\
\verb!MJD-AVG =            51085.979  ! \\
\verb!SPECSYS = `TOPOCENT'! \\
\verb!RESTFRQ =      1.420405752E+09 / [Hz]! \\
\verb!RADESYS = 'FK5'                 ! \\
\verb!  ! \\
\verb!CNAME3Z = 'Barycentric optical velocity'! \\
\verb!RESTWAVZ=          0.211061139 / [m]! \\
\verb!CTYPE3Z = 'VOPT-F2W'            ! \\
\verb!CRVAL3Z =         9.120000E+06  ! \\
\verb!CDELT3Z =       -2.1882651E+04  ! \\
\verb!CRPIX3Z =                 32.0  ! \\
\verb!CUNIT3Z = 'm/s'! \\
\verb!SPECSYSZ= 'BARYCENT'! \\
\verb!SSYSOBSZ= 'TOPOCENT'! \\
\hline
\end{tabular}
\label{ta:header}
\end{table}

\begin{table}
\caption[]{Additional keywords for example header.}
\renewcommand{\arraystretch}{1.05}
\begin{tabular}{l}
\hline
\hline
\noalign{\smallskip}
\verb+123456789 123456789 123456789 123456789 123456789+ \\
\noalign{\smallskip}
\hline
\noalign{\smallskip}
\verb!VELOSYSZ=           26.108E+03 / [m/s]! \\
\verb!  ! \\
\verb!CNAME3F = 'Barycentric frequency'! \\
\verb!CTYPE3F = 'FREQ'                ! \\
\verb!CRVAL3F =    1.37847121643E+09  ! \\
\verb!CDELT3F =         9.764775E+04  ! \\
\verb!CRPIX3F =                 32.0  ! \\
\verb!CUNIT3F = 'Hz'! \\
\verb!SPECSYSF= `BARYCENT'! \\
\verb!SSYSOBSF= 'TOPOCENT'! \\
\verb!VELOSYSF=           26.108E+03 / [m/s]! \\
\verb!  ! \\
\verb!CNAME3W = 'Barycentric wavelength'! \\
\verb!CTYPE3W = 'WAVE-F2W'            ! \\
\verb!CRVAL3W =       0.217481841062  ! \\
\verb!CDELT3W =       -1.5405916E-05  ! \\
\verb!CRPIX3W =                 32.0  ! \\
\verb!CUNIT3W = 'm'! \\
\verb!SPECSYSW= `BARYCENT'! \\
\verb!SSYSOBSW= 'TOPOCENT'! \\
\verb!VELOSYSW=           26.108E+03 / [m/s]! \\
\verb!  ! \\
\verb!RESTFRQR=      1.420405752E+09 / [Hz]! \\
\verb!CNAME3R = 'Barycentric radio velocity'! \\
\verb!CTYPE3R = 'VRAD'                ! \\
\verb!CRVAL3R =    8.85075090419E+06  ! \\
\verb!CDELT3R =       -2.0609645E+04  ! \\
\verb!CRPIX3R =                 32.0  ! \\
\verb!CUNIT3R = 'm/s'! \\
\verb!SPECSYSR= 'BARYCENT'! \\
\verb!SSYSOBSR= 'TOPOCENT'! \\
\verb!VELOSYSR=           26.108E+03 / [m/s]! \\
\verb!  ! \\
\verb!CNAME3V = 'Barycentric apparent radial velocity'! \\
\verb!RESTFRQV=      1.420405752E+09 / [Hz]! \\
\verb!CTYPE3V = 'VELO-F2V'            ! \\
\verb!CRVAL3V =    8.98134229811E+06  ! \\
\verb!CDELT3V =       -2.1217551E+04  ! \\
\verb!CRPIX3V =                 32.0  ! \\
\verb!CUNIT3V = 'm/s'! \\
\verb!SPECSYSV= 'BARYCENT'! \\
\verb!SSYSOBSV= 'TOPOCENT'! \\
\verb!VELOSYSV=           26.108E+03 / [m/s]! \\
\hline
\end{tabular}
\label{ta:ExHeader}
\end{table}

Let us consider an example.  The partial FITS header in
Table~\ref{ta:header} was produced for data from the Very Large
Array. The FITS WCS standard states that all WCS keywords must be
reproduced for the alternate descriptions, but those for non-spectral
axes have been omitted for the sake of brevity and clarity.  We have
also updated the keywords to the standards of Papers I, II, and III\@.
We will derive additional header keywords that could also have been
written.  These are shown in Table~\ref{ta:ExHeader}, again leaving
out axes 1 and 2 of the alternate WCS versions.

The spectral-line cube is regularly sampled in frequency and the
primary description is of a linear frequency axis (from the
\keyw{CTYPE3}\ keyword value \hbox{{\tt 'FREQ'}}).
\keyw{CRVAL3} and \keyw{CRPIX3} record that the 63-channel
spectrum was centered with channel 32 at 1.37835117405\,GHz, some 42
MHz from the rest frequency (\keyw{RESTFRQ} = 1.420405752\,GHz) of
neutral hydrogen (\HI), the line under investigation.  As indicated by
the \keyw{SPECSYS} keyword, it is shown as if it had been observed at
a constant topocentric frequency.  For the moment, let us assume that
the observations were made over a short interval and that this is
correct.

As allowed by Paper I, we use alternate axis description letter
codes in a mnemonic fashion.  Using {\tt Z} for optical, the first
alternate axis description for the spectral axis reflects the
observer's request that a Doppler shift be applied to the \HI\ line
frequency appropriate for a source with a barycentric
optical-convention velocity of 9120\,km\,s$^{-1}$.  Combining
Eqs.~(\ref{eq:P13}) and (\ref{eq:B1}), we get
\begin{displaymath}
 \nu = \nu_0 / (1 + Z/c) ,
\end{displaymath}
\noindent whence the barycentric \HI\ rest frequency shifts to
1.37847122\,GHz, some 120\,kHz greater than the topocentric frequency
given by {\tt CRVAL3}\@.  The difference, of course, simply reflects
the instantaneous relative velocity of the barycentric and topocentric
reference frames at the time of observation.  Using standard software
(the STARLINK program {\tt rv}), with the observatory position defined
by {\tt OBSGEO-X}, {\tt Y}, {\tt Z}, at the time indicated by {\tt
MJD-AVG}, and for the source at (J2000) right ascension and
declination $17^{\rm h}20^{\rm m}26^{\rm s}$,
$-00\degr58\arcmin30\arcsec$ indicated by {\tt CRVAL1} and {\tt
CRVAL2}, it may be verified that the topocentric correction amounts to
26.108\,km\,s$^{-1}$.  Now this correction between reference frames is
a true velocity and thus the relativistic formula, Eq.~(\ref{eq:B3}),
should be used, whereupon the frequency recorded by {\tt CRVAL3} may
be obtained.  The frequency decreases as the observer moves away from
the source position.

With {\tt CRVALZ} set to $9.12 \times 10^6$ m\,s$^{-1}$ it remains to compute
{\tt CDELT3Z}.  This is obtained by transforming {\tt CDELT3} from the
topocentric to the barycentric frame with Eq.~(\ref{eq:B3}) expressed
as
\begin{displaymath}
\Delta\nu = \Delta\nu_0 \frac{c - \varv}{\sqrt{c^2 - \varv^2}} ,
\end{displaymath}
whence the frequency increment becomes 97.64775\,kHz.  Using
Eq.~(\ref{eq:B6}) the corresponding wavelength increment is as
recorded by {\tt CDELT3Z} in Table 14.  While {\tt SPECSYSZ} is set to
{\tt 'BARYCENT'}, {\tt SSYSOBSZ} records the fact that the observation
was actually made from the {\tt 'TOPOCENT'} frame.

The correction for the observatory's motion due to the Earth's
rotation and orbital motion with respect to the barycenter amounted to
26.108\,km\,s$^{-1}$ in the direction of the source.  As illustrated in
Table~\ref{ta:ExHeader}, this velocity shift can be represented in the
FITS header with the new keyword {\tt VELOSYS}\@.

The barycentric frequency may be used as another alternate axis
description.  The reference frequency and frequency increment for {\tt
SPECSYSF = 'BARYCENT'} have already been computed above.  The relevant
keywords have axis description {\tt F} (for frequency) in
Table~\ref{ta:ExHeader}.
   
The frequency description may be translated into a wavelength
description simply using Eqs.~(\ref{eq:B5}) and (\ref{eq:B6}), shown
in Table~\ref{ta:ExHeader} using axis description {\tt W} (for
wavelength).

The frequency description with respect to the {\tt BARYCENT} may
also  be expressed simply as a radio velocity (also regularly sampled)
using Eqs.~(\ref{eq:P3}) and (\ref{eq:P4}).  This is shown as axis
description version {\tt R} (for radio).

A apparent radial velocity description (version {\tt V} for
velocity) requires the use of Eq.~(\ref{eq:B9}) for the reference
value and Eq.~(\ref{eq:B10}) for the increment.

\subsection{Real-world complications}
\label{s:real comp}

An assumption made throughout this paper, particularly in
Sect.~\ref{s:reference}, is that data are provided with at least
celestial coordinate information in addition to the spectral
coordinates discussed here.  The normal method of providing such data
is with the full set of coordinate keywords including {\tt WCSAXES}
with a value of at least 3 (for two celestial and one spectral axis)
even if only one celestial position was observed.  For spectra in
tables, it would be possible to have simple columns for the celestial
coordinates, labeled as such.  But, as discussed at the end of
Sect.~\ref{s:TABtable}, this simple case may still use the full WCS 
nomenclature in order to associate the coordinate columns with the
data column.

One important assumption made in Sect.~\ref{s:S-L_header} was that the
topocentric velocity does not change significantly during the course
of the observation.  We now consider the consequences if this
assumption is violated.  In this we are interested in particular with
three-dimensional ``data cubes'' containing multiple samples on
each of two spatial axes and one spectral axis.

For some instruments, it may be possible to correct the data for each
short exposure to a consistent WCS and then combine exposures to improve
the sensitivity.  For example, single-dish radio telescopes observe a
single direction at a time, allowing the spectral axis in each
observed spectrum to be scaled and resampled before it is added into a
three-dimensional image.
For this type of instrument it is a relatively simple matter to correct
for topocentric motion.

However, other instruments observe multiple directions simultaneously
and also routinely combine multiple exposures to form an image.  While
the error over a single exposure may be negligible, images made by
aperture synthesis radio telescopes, for example, are often based on
multiple exposures taken months apart.  These exposures are usually made
with different array configurations, which may change on a timescale of
months, in order to sample the Fourier space adequately for imaging and
deconvolution.  Therefore, these data are affected not only by the
rotation of the Earth but also by its motion around the Sun which
usually is not negligible.

As a specific example, the actual observation on which
Sect.~\ref{s:S-L_header} is based required an integration time of
several hours.  During this time the observing frequency was adjusted
for the topocentric Doppler shift so that the barycentric frequency at
the map center remained constant.  This ensured that the spectral
feature of interest, associated with a source near the map center,
remained in the same spectral channel.  However, this means that the
primary spectral axis description, based as it is in the topocentric
frame, is not strictly correct.  It was used in this data as an aid to
interferometric analysis, the resulting error being considered
negligible, but strictly speaking a barycentric frequency reference
frame, the {\tt F} alternative WCS in the header, is the correct one.

Even so, the barycentric frame is only strictly correct at the map
center since the topocentric Doppler correction varies across the
field-of-view; the correction is computed as the dot product of the
observatory's velocity vector and a unit direction vector which, of
course, varies across the field.  This gives rise to a differential
error that is greatest when the velocity vector is orthogonal to the
direction to the source and least when it is parallel or anti-parallel.
In other words, the differential correction is greatest when the
correction is zero, and least when the correction is greatest.

For diurnal rotation the worst case differential error occurs for a
telescope at the equator observing a source on the local meridian.  For
an angular offset, $\gamma$, from the tracking center the maximum
differential error is $\vv \sin\gamma$ which amounts to 8.7\,m\,s$^{-1}$
for $\gamma = 1\degr$.  The worst case change in the differential error
over the course of 12 hours occurs for an equatorial telescope
observing a field at the north or south point on the horizon; it is
$2 \vv \sin\gamma$ or 17.5\,m\,s$^{-1}$ for $\gamma =
1\degr$.  If data from exposures separated by several months are
combined then the Earth's 30\,km\,s$^{-1}$ orbital velocity scales 
up the worst-case error to 1.0\,km\,s$^{-1}$ for $\gamma =
1\degr$.  Note that, in the example of Sect.~\ref{s:S-L_header}, the
barycentric correction is close to the maximum value so the
differential correction is nearly at its minimum.

The relevance of the {\tt SSYSOBS}\Ci\ keyword immediately becomes
apparent in this context.  It records the reference frame in which the
frequency has no spatial variation; in other frames the reference
frequency and frequency increment are only strictly correct at the
reference point.  While {\tt SSYSOBS}\Ci\ will most often indicate the
topocentric frame, it is possible to regrid (resample) the spectral data
by applying a position-dependent shift-and-scale so that
{\tt SSYSOBS}\Ci\ has some other value.

Differential Doppler effects are often neglected for contemporary
observations made with small angular fields or modest spectral
resolution.  However, instruments now under development will be capable
of significantly greater sensitivity over wider fields-of-view and/or
spectral resolution and hence will need to consider the effects
discussed here.


\section{Summary} \label{s:summary}

The new keywords required for spectral coordinate systems are summarized
in Table~\ref{ta:Vkeyword}.  {\tt SPECSYS\Ci}, {\tt SSYSOBS\Ci}, {\tt
SSYSSRC\Ci}, and {\tt ZSOURCE\Ci} are used to specify velocity
reference frames; {\tt MJD-AVG}, {\tt DATE-AVG}, {\tt OBSGEO-X}, {\tt
OBSGEO-Y}, {\tt OBSGEO-Z}, {\tt VELOSYS\Ci}, and {\tt ZSOURC\Ci}
enable conversion between these standards of rest; and {\tt
RESTFRQ\Ci}, {\tt RESTFREQ}, and {\tt RESTWAV\Ci} define the spectral
line for which velocities are measured.  Variants of these keywords
for use with tabular data are also defined.

These new keywords and their allowed values, along with new values for
the standard header keywords formalized by Paper I, and the associated
algorithms and methods introduced here allow an accurate description of
spectral coordinates in FITS images.  Wavelength, frequency, apparent
radial velocity and oft-used quantities that are linearly related to
them, such as redshift, energy, and radio velocity, may now be expressed
as functions of pixel coordinate along an axis regularly sampled in
wavelength, frequency, or apparent radial velocity.

A non-linear algorithm is also provided for spectral dispersers commonly
used at optical wavelengths: gratings, prisms, and the combination of
the two, grisms.  Although developed for ideal dispersers, it was shown
to apply quite well to real-world dispersers with suitable fine
adjustment of the instrumental parameters.

The multi-dimensional {\tt -TAB} table lookup algorithm developed for
wavelength calibration will be useful for much more than just spectral
axes.  Several others of the methods introduced here are also completely
general; logarithmic ({\tt -LOG}) coordinates, and the
\keyi{CNAME}{i\Ci} keyword.  The mathematical formalism described in
Sect.~\ref{s:other} for constructing one-dimensional non-linear
coordinate systems also has wide validity.


\appendix
\section{Relativistic space velocities}
\label{appen:relativity}

The discussion of Sect.~\ref{s:basics}, particularly
Fig.~\ref{fig:veltrans}, indicates the importance of transverse, as
well as radial velocities, in spectroscopy at significant velocities.
To be concrete, consider a relativistic jet emanating from a distant
galaxy.  In principle, the galaxy's systemic, cosmological redshift
can be measured separately and used to correct the jet's observed
redshift thereby providing its kinematic redshift (i.e. associated
with a true velocity) in the reference frame of the galaxy. 

Clearly knowledge of the velocity is fundamental in studying jet
kinematics and dynamics.  However, it can only be computed from the
kinematic redshift if the jet's orientation angle is known.  Note that
the equations involving velocity in Table~\ref{ta:speceqn} are
actually only correct if the transverse velocity is zero.  For
example, Eq.~(\ref{eq:B7}) is just a special case of
Eq.~(\ref{eq:Lang}).  This is why the velocity is always referred 
to in the paper as being ``apparent.''

There are instances where the orientation angle may be inferred by
geometry (e.g. by the observed tilt of an accretion disk) or by
modelling.  Defining the orientation angle as $\theta$ we have
\begin{eqnarray}
\vv\sub{r} & = & \vv\sub{s}\, \sin\theta , \\ 
\vv\sub{t} & = & \vv\sub{s}\, \cos\theta , \\
\vv\sub{s} & = & \sqrt{\vv\sub{r}^2 + \vv\sub{t}^2}  ,
\end{eqnarray}
where $\vv\sub{r}$ is the radial velocity, $\vv\sub{t}$ is the transverse
velocity, and $\vv\sub{s} \ge 0$ is the total space velocity.  The
orientation angle $\theta$ is defined so that $\theta = -90\degr$ is
towards the observer, $\theta = 0$ is transverse, and $\theta = +90\degr$
is away from the observer.  It is beyond the scope of this paper to
describe how $\theta$ should be determined or what exactly it means
for particular relativistic observers, other than that it resolves the
kinematic velocity into radial and tangential components.  With these
definitions, the equations of Table~\ref{ta:speceqn} are given in
Table~\ref{ta:vvtheta}.  Note that for $\chi > 1$
Eqs.~(\ref{eq:vspace}) and (\ref{eq:dvsdchi}) have two valid solutions
for $\theta < 0$ but none otherwise.  For $\chi \le 1$ they have only
one valid solution for any value of $\theta$.  Invalid solutions,
i.e.\ those with $\vv\sub{s} < 0$, are the negative of the valid
solution for $-\theta$.

\begin{table}
\caption[]{Velocity equations using orientation and total velocity.}
\vspace{-20pt}
\begin{eqnarray}
\hline\hline
\noalign{\smallskip}
\nu & = & \nu_0
   \frac{\sqrt{c^2 - \vv\sub{s}^2}}
        {c + \vv\sub{s}\sin\theta} , \\
\Dfrac{\nu}{\vv\sub{s}} & = & -c\nu_0
   \frac{\vv\sub{s}+c\sin\theta}
        {(c + \vv\sub{s}\sin\theta)^2 \sqrt{c^2 - \vv\sub{s}^2}} , \\
\lambda & = & \lambda_0
   \frac{c + \vv\sub{s}\sin\theta}
        {\sqrt{c^2 - \vv\sub{s}^2}} , \\
\Dfrac{\lambda}{\vv\sub{s}} & = & c\lambda_0
   \frac{\vv\sub{s} + c\sin\theta}
        {(c^2 - \vv\sub{s}^2)^{3/2}} , \\
\vv\sub{s} & = & c 
   \frac{-\chi^2\sin\theta \pm \psi}
        {\chi^2 + \psi^2} , \label{eq:vspace} \\
{\rm where} && \chi = \frac{\nu}{\nu_0} = \frac{\lambda_0}{\lambda} , \\
{\rm and}   && \psi = \sqrt{1 - \chi^2\cos^2\theta} , \\
\noalign{\smallskip}
\Dfrac{\vv\sub{s}}{\nu} & = & \frac{1}{\nu_0} \Dfrac{\vv\sub{s}}{\chi} , \\
\Dfrac{\vv\sub{s}}{\lambda} & = & \frac{-\lambda_0}{\lambda^2}
   \Dfrac{\vv\sub{s}}{\chi} , \\
{\rm where} && \Dfrac{\vv\sub{s}}{\chi} = -c\chi
   \frac{2\psi\sin\theta \pm (1 + \psi^2\sin^2\theta)}
        {\psi(\chi^2 + \psi^2)^2} \cdot \label{eq:dvsdchi} \\
\noalign{\smallskip}
\hline\nonumber
\end{eqnarray}
\label{ta:vvtheta}
\vspace{-20pt}
\end{table}

There are other combinations that are of interest, including
\begin{eqnarray*}
\sin\theta & = &\frac{\nu_0\sqrt{c^2-\vv\sub{s}^2}}{\nu\vv\sub{s}} -
                \frac{c}{\vv\sub{s}} , \\
   & = &\frac{\lambda\sqrt{c^2-\vv\sub{s}^2}}{\lambda_0\vv\sub{s}} -
                \frac{c}{\vv\sub{s}} \cdot \\
\end{eqnarray*}

It would perhaps be more instructive to express these sorts of
equations in terms of the actual radial and transverse velocities.
This is relatively straightforward, but the equations become even more
messy and so we omit them here.

It has been proposed that we introduce a single keyword to express
$\theta$ so that we may express velocities in terms of a true space
velocity $\vv\sub{s}$ rather than an apparent velocity $\vv$ as done
throughout this paper.  However, astrophysical jets are thought to be
conical in form, to interact along their boundaries with the external
medium, and to change direction, even assuming helical shapes.  Thus,
there is no single angle $\theta\/$ that can normally be used to
describe all directions in an image and one should probably not even
assume that $\theta$ is independent of redshift in any one direction.
For those cases in which a single angle may be appropriate, such as a
spectrum of a spatially limited region in a relativistic jet, we
reserve the keyword
\begin{center}
\begin{tabular}{l}
\noalign{\vspace{-5pt}}
\keyw{VELANGL\Ci} \hspace{2em} (floating-valued)\\
\noalign{\vspace{-5pt}}
\end{tabular}
\end{center}
\noindent with default value $+90\degr$ to express the angle $\theta$
in degrees.  All of this discussion has neglected non-velocity
redshifts such as those due to the gravity of the black hole thought
to be at the base of the astrophysical jet.  Therefore, we choose to
leave the full expression of internal velocities within a celestial
object to those observers who are armed with both extensive
observations and a detailed model.  Apparent velocity is a suitable,
semi-physical concept appropriate to a general FITS standard.


\begin{acknowledgements}

The authors wish to acknowledge constructive comments and
contributions from the following:  Wim Brouw (Australia Telescope
National Facility), Bob~Hanisch (Space Telescope Science Institute), 
Jonathan McDowell (Harvard University), William Pence (Goddard Space
Flight Center), Arnold Rots (CfA Harvard University),  William
Thompson (Goddard Space Flight Center), Doug~Tody (National Radio
Astronomy Observatory), and Patrick Wallace (U.K. Starlink).

The {\tt SLALIB} software library developed by Patrick Wallace was
helpful in the preparation of  Sect.~\ref{s:reference} and should be
of considerable help to users of conventions described in this
work.  It is available via {\tt http://www.starlink.ac.uk/}.

The National Radio Astronomy Observatory is a facility of the (U.S.)
National Science Foundation operated under cooperative agreement by
Associated Universities, Inc. 

The Australia Telescope is funded by the Commonwealth of Australia for
operation as a National Facility managed by CSIRO\@.

The National Optical Astronomy Observatory is a facility of the (U.S.)
National Science Foundation operated under cooperative agreement by
Associated Universities for Research in Astronomy, Inc.

UCO/Lick Observatory is operated by the University of California.
\end{acknowledgements}



\end{document}